\documentclass[useAMS,usenatbib]{mn2e}
\usepackage{color}
\usepackage{graphicx}
\usepackage{amssymb}
\usepackage{array}
\usepackage{color}
\usepackage{lscape}
\usepackage{natbib}
\bibpunct{(}{)}{;}{a}{,}{,}

\newcommand{\degree}{\ensuremath{^\circ}}
\title[Probing the magnetic fields in L1415 and L1389]{Probing the magnetic fields in L1415 and L1389}
\author[Soam et~al.]{A. Soam,$^{1,2}$\thanks{email:{archana@kasi.re.kr, archanasoam.bhu@gmail.com}}
Chang Won Lee,$^{1,3}$ Maheswar G.,$^{2}$ Gwanjeong Kim,$^{1,3}$Neha S.,$^{2,4}$
\newauthor % starts a new line in the
Mi-Ryang Kim$^{1,5}$
\\
$^{1}$ Korea Astronomy \& Space Science Institute (KASI), 776 Daedeokdae-ro, Yuseong-gu, Daejeon 305-348, Republic of Korea.\\
$^{2}$ Aryabhatta Research Institute of Observational Sciences (ARIES), Nainital 263002, India.\\
$^{3}$ University of Science \& Technology, 176 Gajeong-dong, Yuseong-gu, Daejeon, Republic of Korea.\\ 
$^{4}$ Pt. Ravishankar Shukla University, Amanaka G.E.Road, Raipur, Chhatisgarh, India - 492010\\
$^{5}$ Chungbuk National University, Chungdae-ro 1, Seowon-gu, Cheongju, Chungbuk 28644, Republic of Korea}

\begin{document}
\date{Submitted------}

\pagerange{\pageref{firstpage}--\pageref{lastpage}} \pubyear{--}

\maketitle

\label{firstpage}

\begin{abstract}
We present the R-band polarimetric results towards two nebulae L1415 and L1389 containing low luminosity stars. Aim of this study is to understand the role played by magnetic fields in formation of low luminosity objects. Linear polarization arise due to dichroism of the background starlight projected on the cloud providing the plane-of-the sky magnetic field orientation. The offsets between mean magnetic field directions obtained towards L1415 and L1389 and the projected outflow axes are found to be 35$^{\degree}$ and 12$^{\degree}$, respectively. The offset between cloud minor axes and mean envelope magnetic field direction in L1415 and L1389 are 50$^{\degree}$ and 87$^{\degree}$, respectively. To estimate the magnetic field strength by using the updated Chandrasekhar-Fermi relation, we obtained the $^{12}$CO(J=1-0) line velocity dispersion value towards L1415 cloud using the TRAO single dish observations. The values of B$_{pos}$ in L1415 and L1389 are found to be 28$~\mu$G and 149$~\mu$G using CF technique and 23$~\mu$G and 140$~\mu$G using structure function analysis, respectively. The values of B$_{pos}$ in these clouds are found to be consistent using both the techniques. By combining the present results with those obtained from our previous study of magnetic fields in cores with VeLLOs, we attempt to improve the sample of cores with low luminosity protostars and bridge the gap between the understanding of importance of magnetic fields in cores with VeLLOs and low luminosity protostars.  The Results of this work and that of our previous work show that the outflow directions are aligned with envelope magnetic fields of the clouds.
\end{abstract}

%#############################             INTRODUCTION

\section{Introduction}

Inspite of the major progress in understanding the process of low mass star formation in the past three decades \citep[e.g. volumes by ][]{1993JBAA..103..266L}, the earliest stages of the low mass star formation remains poorly known. According to the present idea of low mass star formation \citep{1987ARA&A..25...23S} in isolated cores, the gravitationally bound dense starless cores are the first evolutionary phase in the path of molecular clouds to the formation of stars. It is believed that the key ingredients involved in the star formation are gravity, magnetic fields and turbulence. The influence of magnetic field on various stages of star formation is not clearly known.  In the magnetic field dominated scenario, the prestellar cores \citep[cores on the verge of star formation;][]{1994MNRAS.268..276W} are thought to be initially supported against their self-gravity by magnetic fields and evolve to higher central condensation through Ambipolar-diffusion \citep{1991ApJ...373..169M}. In the magnetically dominated scenario, the material is settled in a disc like structure. This disc like structure is of few thousands AU in size. This mediation of material by the field lines can later result into the magnetic fields parallel to the symmetry axis of the disc like structure. A pinched hour glass morphology of field lines are seen in the inner core region within the infall radius \citep{1993ApJ...415..680F, 1993ApJ...417..243G}. The field lines in the envelope can be moderately pinched and join smoothly with the inner field lines. But in the cores where magnetic field play relatively lesser role, the available studies \citep{2004RvMP...76..125M,2007ApJ...661..262D, 2010ApJ...723..425D} suggest that the clumps and subsequent cores are formed on the junctions of the turbulent flows with supersonic motions. The supersonic turbulent flows can randomize the weaker magnetic field geometry \citep{2004Ap&SS.292..225C}. Now these models can/should be tested by mapping the magnetic fields on the different spatial scales of the clouds and cores. Such studies can help in finding the correlations (if present) in the various cloud properties such as minor axes, bipolar outflows and kinematics with the magnetic field morphology.

The plane-of-the-sky magnetic field maps of dense cores are produced using polarization measurements in near infrared and optical wavelengths caused due to the selective extinction of background starlight \citep{1977ApJ...215..529D, 2003JQSRT..79..881L}. The preferential extinction of the background starlight is caused by the presence of elongated dust grains which are aligned with their minor axes parallel to the ambient magnetic fields. The actual mechanism by which these dust grains align with the magnetic fields is still under debate \citep{2007JQSRT.106..225L, 2015ARA&A..53..501A}. However, radiative torque mechanism, first proposed by \citet{1976Ap&SS..43..291D}, seems to be emerging as the most successful one in explaining the dust grain alignment in various environments \citep[e.g., ][]{2014MNRAS.438..680H, 2015MNRAS.448.1178H, 2015ARA&A..53..501A}. 

We have chosen a poorly studied cloud L1415 for our present work. This cloud is classified as  opacity class 3 and consist of a low luminosity protostar ($0.13~L_{\odot}$) namely IRAS04376+5413 \citep{2007A&A...463..621S}. In the present study, we are showing the R-band polarization results towards L1415 region with an aim to map and study the plane-of-the-sky magnetic field morphology in the low density regions of this cloud. We made $^{12}$CO(J=1-0) molecular line observations of this cloud to estimate the velocity dispersion in the cloud which is required to estimate the magnetic field strength. 
 
 For our study, we also selected another cloud L1389, a cometary-shaped cloud \citep{2013A&A...551A..98L} with a round head (consisting of IRAS source) and sharp tail. This cloud is located nearby Perseus and is associated with the Lindbald ring \citep{1973A&A....24..309L}. Herschel-SPIRE images of L1389 clearly show a cometary morphology of the cloud \citep{2013A&A...551A..98L}. Based on the results of available infrared (IR) and (sub)millimeter continuum observations, \citet{2010ApJS..188..139L} found two sources present in L1389 cloud head. One source is found as L1389-IRS and dominates the infrared emission in the Spitzer images but it has a faint millimeter continuum emission. But the other source is referred as L1389-SMM and it has no IR emission in Spitzer. On the contrary, this source is found to be with millimeter continuum emission. The spectral energy distributions (SEDs) fitting results from \citet{2012ApJ...751...89C} indicated that L1389-IRS may be a Class 0/I transition object while L1389-SMM (also referred as L1389-MMS) may be a prestellar core but later when they compared this source with prestellar cores and Class 0 protostars and found that L1389-MMS is more evolved than prestellar cores but less evolved than Class 0 protostars. They also found that the observed characteristics of this source are similar to the theoretical predictions from radiative/magnetohydrodynamical simulations of first hydrostatic core \citep{1969MNRAS.145..271L, 1998ApJ...495..346M}. But they also suggested that this source may be a possible extremely low luminosity star embedded in an edge-on disc. L1389-IRS is directly observed in the near-IR wavelengths \citep{2010ApJS..188..139L} which means it can be a Class 0/I transition object with a luminosity of $L_{bol}$ $\sim0.5~L_{\odot}$. The estimated luminosity of L1389-MMS is found to be less than $0.04~L_{\odot}$ \citep{2012ApJ...751...89C}.
 
This paper is organized such that section \ref{observe} describes the methods of data acquisition and reduction procedure using optical polarimetric technique and $^{12}$CO(1-0) molecular line observations. In section \ref{result}, we present our results of optical polarization and molecular line observations. In section \ref{discuss}, we discuss the procedure used to subtract the foreground polarization contribution. We also discuss the magnetic field geometry and strength in the same section. Finally, we conclude our paper by summarizing the results in section \ref{conclude}.

%==============
%\input{table1_log}		%Observation Log
\begin{table}
\begin{small}
  \caption{Log of optical polarization \& radio observations.}\label{tab:obslog}
  \begin{tabular}{llll}
\hline\hline
            \multicolumn{3}{c}{Polarization observation}\\
\hline
Cloud          &  Observation date (year, month,date) & Band\\
\hline
 L1415              & 2011 Nov. 23 \& 26; 2011 Dec 19, 20 \& 24     &  $R_c$            \\  
                         & 2013 Oct. 29         &              \\  
 L1389              & 2011 Nov. 26; 2013 Oct. 28, 29     &  $R_c$            \\ 
 \hline
             \multicolumn{3}{c}{Radio observation}\\
\hline
Cloud          &  Observation date (year, month,date) & Freq.\\
L1415             & 2016 March 4        &              115 GHz  \\
\hline\hline 
\end{tabular}
\end{small}
\end{table}
%==============

\section{OBSERVATIONS AND DATA REDUCTION}\label{observe}
\subsection{Optical polarization}

The log of the R-band polarimetric observations of 8 fields towards L1415 and 4 fields towards L1389 are shown in Table \ref{tab:obslog}.  We made these observation from Aries IMaging POLarimeter i.e. AIMPOL \citep{2004BASI...32..159R}. This polarimeter is a back-end instrument on 1 m diameter optical telescope at Aryabhatta Research Institute of Observational Sciences (ARIES), India. AIMPOL consists of an achromatic half-wave plate (HWP) modulator and a Wollaston prism beam-splitter. The images were obtained with the use of 1024$\times$1024 pixel$^{2}$ CCD chip (Tektronix;  TK1024) out of which central 325$\times$325 pixel$^{2}$ area is used for the imaging because of the $\sim 8^{\prime}$ diameter field of view of the CCD with plate scale 1.48$^{\prime\prime}/$pixel. The stellar image size falls in 2 to 3 pixels. The gain and read out noise of CCD are 11.98 $e^{-}$  and 7.0 $e^{-}$ per ADU, respectively. 

The imaging polarimeters use a polarization modulator followed by an analyser to convert any polarized component into a light intensity. Fluxes of ordinary ({\it $I_{o}$}) and extraordinary ({\it $I_{e}$}) beams for all the observed sources with a good signal to noise ratio were extracted by standard aperture photometry using the IRAF package. The ratio {\it {R($\alpha$)}} is obtained as

 \begin{equation}
 R(\alpha) = \frac{\frac{{I_{e}}(\alpha)}{{I_{o}}(\alpha)}-1} {\frac{I_{e}(\alpha)} {I_{o}(\alpha)}+1} =  P cos(2\theta - 4\alpha), 
\end{equation}
where {\it $P$} is the degree of polarization and $\theta$ is the position angle of the plane of polarization. Here {\it $\alpha$} is the orientation of the fast axis of HWP at $0^{\circ}$, $22.5^{\circ}$, $45^{\circ}$ and $67.5^{\circ}$ corresponding to four normalized Stokes parameters, respectively, q[R($0^{\circ}$)], u[R($22.5^{\circ}$)], $q_{1}$[R($45^{\circ}$)] and $u_{1}$[R($67.5^{\circ}$)]. We estimated the errors in normalized Stokes parameters ($\sigma_R$)($\alpha$)($\sigma_q$, $\sigma_u$, $\sigma_{q1}$, $\sigma_{u1}$) using the relation provided by \citet{1998A&AS..128..369R}.

The instrumental polarization of AIMPOL has been checked by observing un-polarized standards and found nearly invariable with a value of $\sim0.1\%$ by \citet{2013MNRAS.432.1502S, 2015A&A...573A..34S} and \citet{2016A&A...588A..45N}. The polarized standard stars observed by us to calibrate our polarization values are summarized in Table \ref{tab:std}.

%************Std table*************
%\input{table2_pol_std}
\begin{table}
\begin{small}
\caption{Polarized standard stars observed in $R_c$ band.}\label{tab:std}
\begin{tabular}{lll}\hline
Date of     &P $\pm$ $\epsilon_P$ 	&  $\theta$ $\pm$ $\epsilon_{\theta}$  \\
Obs.		&(\%)            		& ($\degree$)                           \\\hline
\multicolumn{3}{l}{{\bf HD$~$236633} ($^\dagger$Standard values: 5.38 $\pm$ 0.02\%, 93.04 $\pm$ 0.15$\degree$)}\\
26 Nov 2011	& 5.3 $\pm$ 0.1     & 92 $\pm$ 1 \\
19 Dec 2011 & 5.5 $\pm$ 0.1     & 93 $\pm$ 1 \\
20 Dec 2011	& 5.3 $\pm$ 0.2     & 93 $\pm$ 2 \\
24 Dec 2011	& 5.7 $\pm$ 0.2     & 93 $\pm$ 1 \\\hline
\multicolumn{3}{l}{{\bf HD$~$236954} ($^\ddagger$Standard values: 5.79 $\pm$ 0.09\%, 111.20 $\pm$ 0.49$\degree$)}\\
20 Dec 2011 & 6.0 $\pm$ 0.1     & 111 $\pm$ 2\\\hline
\multicolumn{3}{l}{{\bf BD$+$59$\degree$389} ($^\dagger$Standard values: 6.43 $\pm$ 0.02\%, 98.14 $\pm$ 0.10$\degree$)}\\
26 Nov 2011 & 7.0 $\pm$ 0.2     & 98 $\pm$ 1  \\
19 Dec 2011 & 7.7 $\pm$ 0.2     & 99 $\pm$ 1  \\
20 Dec 2011 & 6.2 $\pm$ 0.1     & 98 $\pm$ 1  \\
24 Dec 2011 & 6.4 $\pm$ 0.1     & 98 $\pm$ 1 \\\hline
\multicolumn{3}{l}{{\bf HD$~$204827} ($^\dagger$Standard values: 4.89 $\pm$ 0.03\%, 59.10 $\pm$ 0.17$\degree$)}\\
20 Oct 2013 & 5.0 $\pm$ 0.2 	& 66 $\pm$ 7 \\        \hline
\multicolumn{3}{l}{{\bf HD$~$19820} ($^\dagger$Standard values: 4.53 $\pm$ 0.02\%, 114.46$\pm$ 0.16$\degree$)}\\
20 Oct 2013 & 4.5 $\pm$ 0.1     & 116 $\pm$ 1  \\\hline
\end{tabular}

$\dagger$  Values in R band from \citet{1992AJ....104.1563S} \\
$\ddagger$ Values in V band from \citet{1992AJ....104.1563S} \\
\end{small}
\end{table}
%*********************************

\subsection{Radio Observations}
We carried out the On-The-Fly (OTF) mapping observations of a region of $25\arcmin\times25\arcmin$ around L1415-IRS in $^{12}$CO(1-0) and C$^{18}$O(1-0) molecular line simultaneously on March 04, 2016 using the instrument SEcond QUabbin Observatory Imaging Array (SEQUOIA) array at Taeduk Radio Astronomical Observatory (TRAO). TRAO is a millimeter-wave radio observation facility with a single dish of 13.7m diameter at Korea Astronomy and Space Science Institute (KASI) in Daejeon, South Korea. SEQUOIA-TRAO is equipped with high-performing 16 pixel MMIC preamplifiers in a 4$\times$4 array, operating within 85$\sim$115 GHz frequency range. The system temperature is ranging from 250 K (86$\sim$110 GHz) to 500 K (115 GHz; $^{12}$CO). As the optical system provides 2-sideband, two different lines can be observed simultaneously. The sky signals were subtracted in position switch mode. The beam size (HPBW) and main beam efficiency of the telescope are about $44^{''}$ and 54$\pm$2\% at 115 GHz, respectively (TRAO staff private communication). The integration time of OTF mapping was $\sim$180 minutes to achieve a rms of 0.3 K in ${T_{A}}^{\ast}$ scale in both the line. The signal-to-noise ratio is measured to be $\sim$16 at a brightest position of ${T_{A}}^{\ast}\sim 4.8$ K. The achieved velocity resolution is $\sim 0.2$ km $s^{-1}$. The pointing and focus of the telescope was done using the source Orion A in SiO line. The pointing of the telescope was as good as $\sim~5-7^{''}$ and the system temperature was within 500-650 K during the observations. The data were reduced by the CLASS software of the GILDAS package and further analysis was done using Common Astronomy Software Applications (CASA) and python language.

%***************************  RESULTS  ******************************************* 

\section{Results}\label{result}
Results of our optical polarization measurements of stars projected on L1415 and L1389 are given in Table \ref{tab:polresults}. The columns of the table give star number, corresponding right ascension (J2000) and the declination (J2000), the degree of polarization (P) in percent (\%) and polarization angle ($\theta_{P}$) in degree, respectively. The values of $\theta_{P}$ are measured from north increasing towards east. The data points with ratio of P and the error in P ($\sigma_{P}$), $\frac{P}{\sigma_{P}}>$2 are considered in this study.

Several reactions forming different molecules take place on the surface of the dust grains in the molecular clouds at their different evolutionary stages. These reaction can change the characteristics such as shape, size and composition of the grains. The grains located inside the denser parts of the molecular clouds are found to be bigger than the grains on the outer periphery \citep{1980ApJ...235..905W, 2003AJ....126.1888K, 2005ASPC..343..321W, 2010A&A...522A..84O}. The dust grains with the similar size as those on the outer regions of the clouds efficiently polarize the light in the optical wavelengths \citep{1996ASPC...97..325G}. Therefore, the magnetic field maps made in this study are in the lower density outer envelope regions of L1415 and L1389.

\subsection{L1415}

\begin{figure}
\resizebox{8cm}{13cm}{\includegraphics{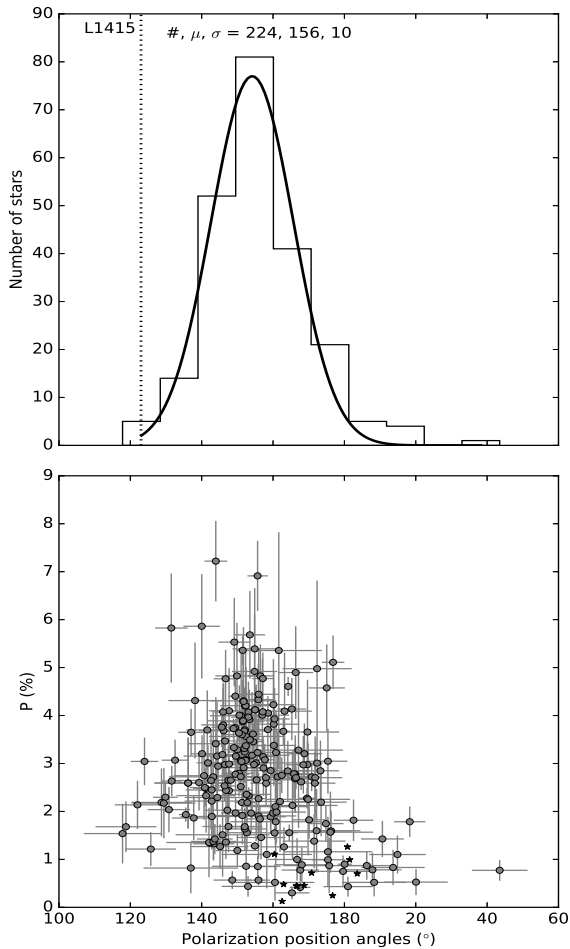}}
\caption{{\bf Upper panel}: Gaussian fitted histogram of $\theta_{P}$ with bin size $10\degree$ in L1415. The orientation of Galactic plane is drawn using dotted line. {\bf Lower panel}: Variation of P with $\theta_{P}$ of program stars and that of the stars which are foreground to the cloud L1415. Filled circles represent program stars and foreground stars are shown by filled star symbols}\label{Fig:correctGaussL1415}
\end{figure}

\begin{figure}
\resizebox{8.4cm}{8.1cm}{\includegraphics{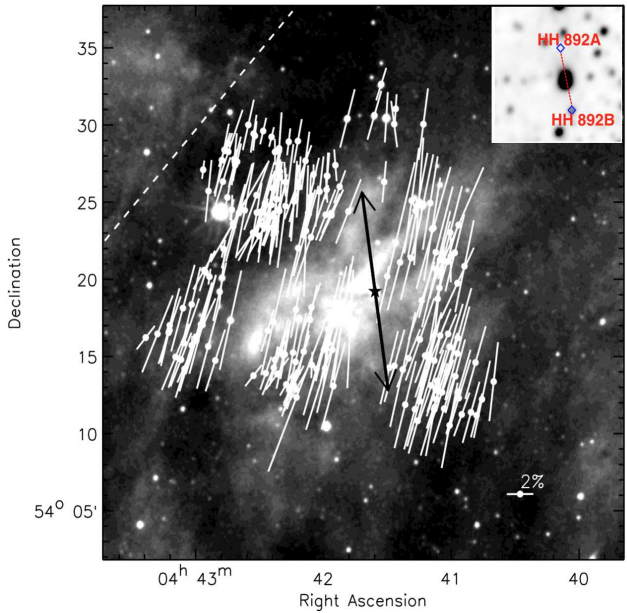}}
\caption{Overlaid polarization vectors of the stars projected on L1415 shown on $0.6^\circ$ $\times$ $0.6^\circ$ WISE 12$~\mu$m image after subtracting the foreground polarization contribution. The double arrow headed line shows the direction of CO outflow from L1415-IRS (shown by star symbol) and the dashed line shows the orientation of the Galactic plane. A vector with 2$\%$ polarization is shown as reference. The inset shows the positions of HH objects \citep{2007A&A...463..621S} on $0.04^\circ$ $\times$ $0.04^\circ$ WISE 3.4$~\mu$m image of L1415.}\label{Fig:correctL1415}
\end{figure}

The polarization measurements for 224 stars observed towards dark nebula L1415 in R-band are carried out in this work. The upper panel of Fig. \ref{Fig:correctGaussL1415} shows the Gaussian fitted histogram of $\theta_{P}$ of the stars towards L1415. The lower panel of this figure shows the P versus $\theta_{P}$ plot of these stars. In Fig. \ref{Fig:correctL1415}, we show the optical polarization vectors overlaid on $0.6^\circ$ $\times$ $0.6^\circ$ WISE 12$~\mu$m image of L1415. Here the lengths of polarization vectors correspond to the degree of polarization. A vector with 2\% polarization is shown in the figure as a reference vector. The  dashed-line shows the direction of Galactic plane with a plane of sky projection angle of 123$^\circ$. The inset shows the positions of Herbig Haro objects \citep[HHOs;][]{2007A&A...463..621S} on $0.04^\circ$ $\times$ $0.04^\circ$ WISE 3.4$~\mu$m image of L1415. The mean values of P and $\theta_{P}$ with their corresponding standard deviation values are found to be 3.1$\pm$1.3\% and 155$\pm$10$\degree$, respectively. Among the targets we observed, there are no peculiar type stars and young stellar objects hence our polarization values can be used to estimate the local magnetic field properties of L1415. There are no near-IR and submm polarization studies available towards L1415. 

The evidence of well collimated outflow is found to be associated to L1415-IRS, with the presence of HH 892 \citep{2007A&A...463..621S}. The spectrum of HH 892A resembles to that of a high excitation HHO \citep{1996MNRAS.280..567R}. An active Herbig-Haro flow towards L1415 has also been reported by \citet{2007A&A...463..621S} based on the long-slit spectroscopic study. On the DSS image of L1415, a HHO candidate near IRAS 04376+5413 was detected \citep{2007A&A...463..621S}. Using subsequent H$_{\alpha}$ and [SII] imaging, \citet{2007A&A...463..621S} confirmed the presence of the emission line object with no counterpart. This HHO was assigned as 892 in Reipurth's catalog \citep{1999hhoe.conf.....R}. The NIR morphology of L1415-IRS also shows a bipolar nebulosity, with two lobes seen in 2.16$~\mu$m \citep{2007A&A...463..621S}.

\begin{figure}
\resizebox{8.5cm}{7.6cm}{\includegraphics{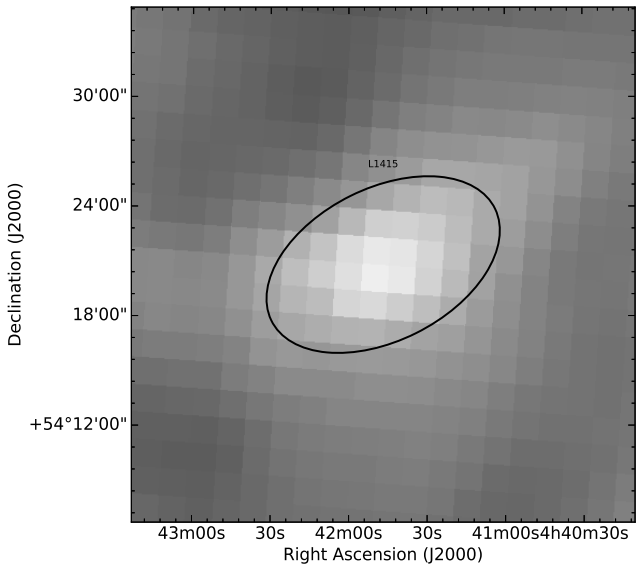}}
\caption{Ellipses fitted to the IRAS 100$~\mu$m image of L1415 for estimating the minor axis position angle using the CSAR based on CLUMPFIND.}\label{Fig:L1415_IRAS100}
\end{figure}

No prior studies have given any information of the major and minor axes of the cloud. We estimated the minor axis position angle of L1415 core by fitting an ellipse in IRAS 100$~\mu$m data (no Herschel and SCUBA\footnote{Submillimetre Common-User Bolometer Array on James Clark Maxwell Telescope} data are available for this cloud). We performed the photometry on the IRAS cores using the CSAR (Cardiff Source-finding AlgoRithm) based on CLUMPFIND. The CSAR is designed to find prestellar/starless clumps with the sizes in the range of $\sim$5000 AU$-$2.7 pc by removing background emission where the dense cores are embedded \citep{2013MNRAS.432.1424K}. We calculated the background by averaging the pixel values outside the region over which the photometry is measured. Fig. \ref{Fig:L1415_IRAS100} shows the ellipse fitted to the IRAS 100$~\mu$m image of L1415. CSAR works by sorting each pixel in the map with respect to signal-to-noise ratio in order of decreasing intensity. The ellipse is fitted at a level where the intensity value is basically the FWHM of the central pixel value and then falls gradually. The minor axis position angle of the fitted ellipse is found to be $\sim$25$\degree$ which is the projection on the sky. We adopted this position angle for our further analysis of the results in L1415 cloud.

%----------------------------Horizontal layout--------------
\begin{figure*}%[t]
\includegraphics[width=85mm]{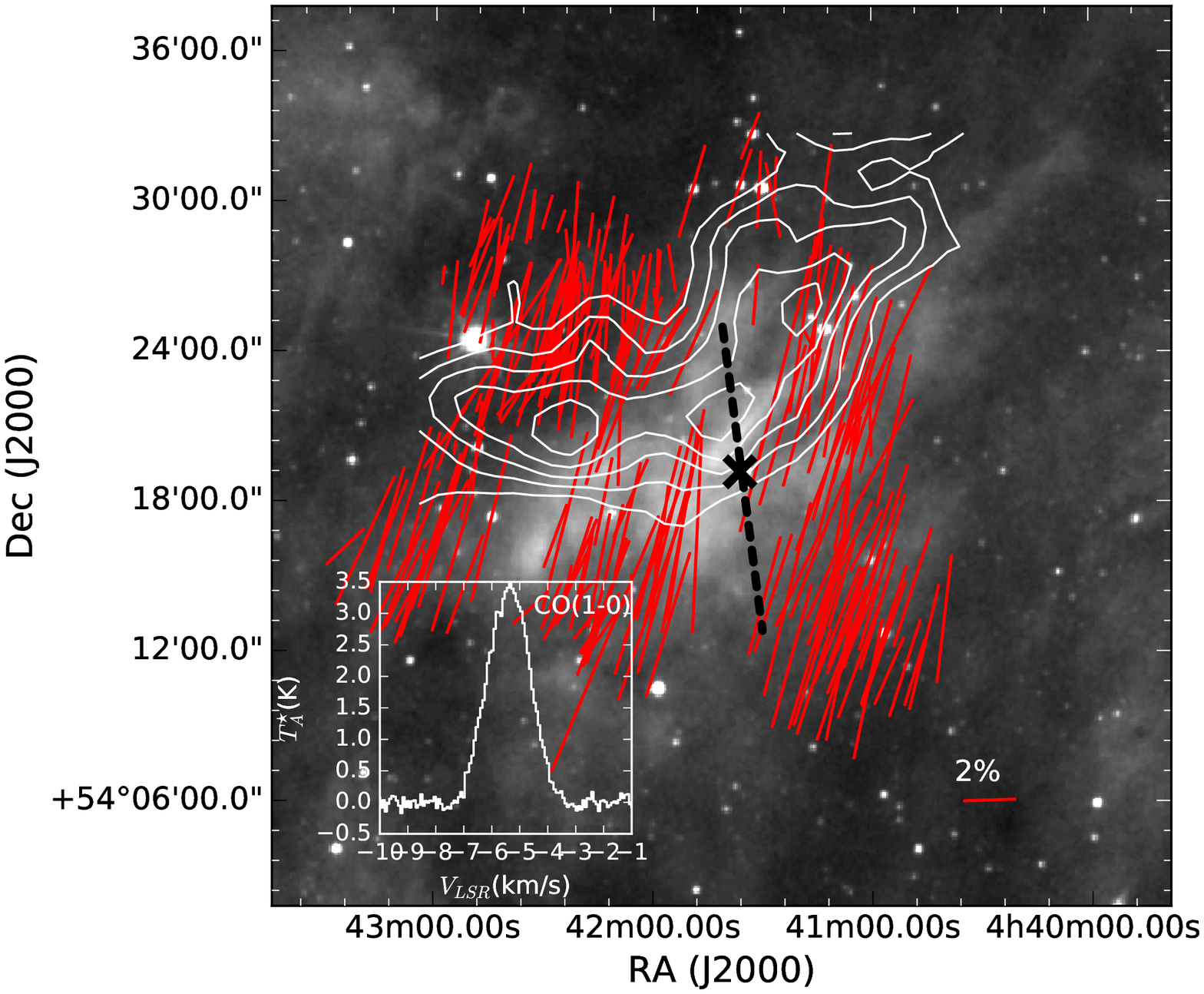}%
\includegraphics[width=85mm]{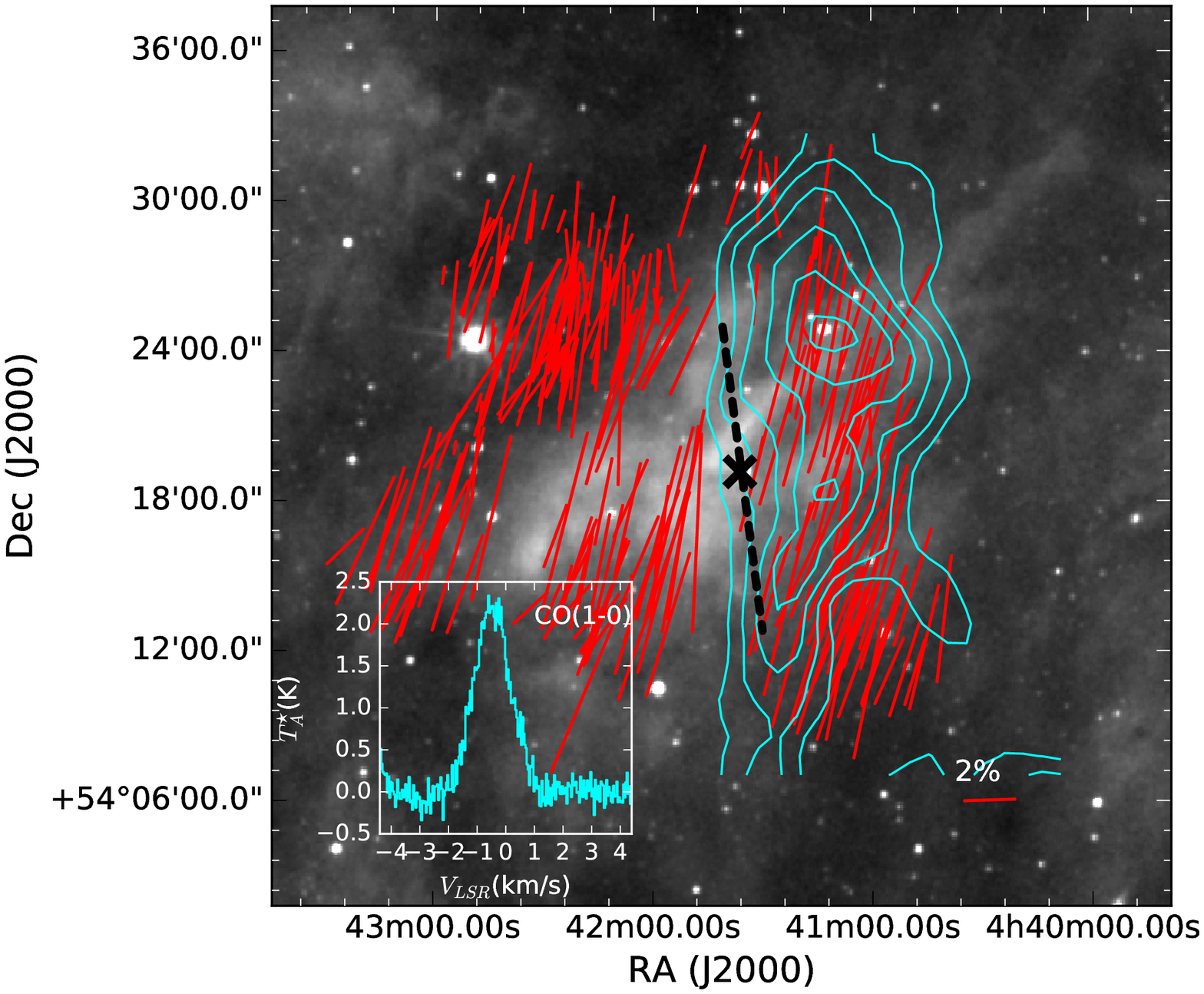}%
\caption{WISE 12$~\mu$m image of L1415 with overplotted $^{12}$CO(1-0) integrated intensity contours corresponding to two velocity components with $V_{LSR}$ -5.5$~$km$~s^{-1}$ (white in left panel) and -0.5$~$km$~s^{-1}$ (cyan in right panel) overlaid with optical polarization vectors in red color. The contour levels for the integrated intensity in left panel are drawn between 7.2$~$K$~$km$~s^{-1}$ and 14.6 $~$K$~$km$~s^{-1}$ with an increment of 1.2 $~$K$~$km$~s^{-1}$. The contour levels for the integrated intensity in right panel are drawn between 63$~$K$~$km$~s^{-1}$ and 188 $~$K$~$km$~s^{-1}$ with an increment of 9$~$K$~$km$~s^{-1}$. The average spectra corresponding to the emissions are shown in the insets in both the panels. The positions of L1415-IRS and the outflow directions in both the panels are shown using black cross and the dashed line, respectively.}\label{Fig:l1415_CO_C18O}
\end{figure*}
%---------------------------------------------------------------

\subsection{Molecular line results in L1415}
Fig. \ref{Fig:l1415_CO_C18O} shows WISE 12$~\mu$m image of L1415 with overplotted $^{12}$CO(1-0) integrated intensity contours corresponding to two velocity components with $V_{LSR}$ -5.5$~$km$~s^{-1}$ (white in left panel) and -0.5$~$km$~s^{-1}$ (cyan in right panel) found in the direction of L1415. The average spectra corresponding to the emissions are shown in the insets in both the panels. The images are also overlaid with optical polarization vectors in red color. The positions of L1415-IRS and the outflow directions in both the panels are shown using black cross and the dashed line, respectively. The width of spectral line is generally expressed as FWHM and often the line profiles are well represented by Gaussian-shape. From the OTF map in $^{12}$CO(1-0) molecular line towards L1415, we estimated the $^{12}$CO line width as it depicts the low density region of the cloud where optical polarization observations are made. The average value of the $^{12}$CO(1-0) line widths measured at various positions of the emission with $V_{LSR}$ -5.5$~$km$~s^{-1}$ is found to be 1.65$\pm$0.02$~$km$~s^{-1}$ and that of the other component with $V_{LSR}$ -0.5$~$km$~s^{-1}$ is found to be 1.64$\pm$0.02$~$km$~s^{-1}$. The width measurement in the spectra is done by fitting Gaussian to the lines using CLASS software. 

\begin{figure}
\resizebox{8.0cm}{14cm}{\includegraphics{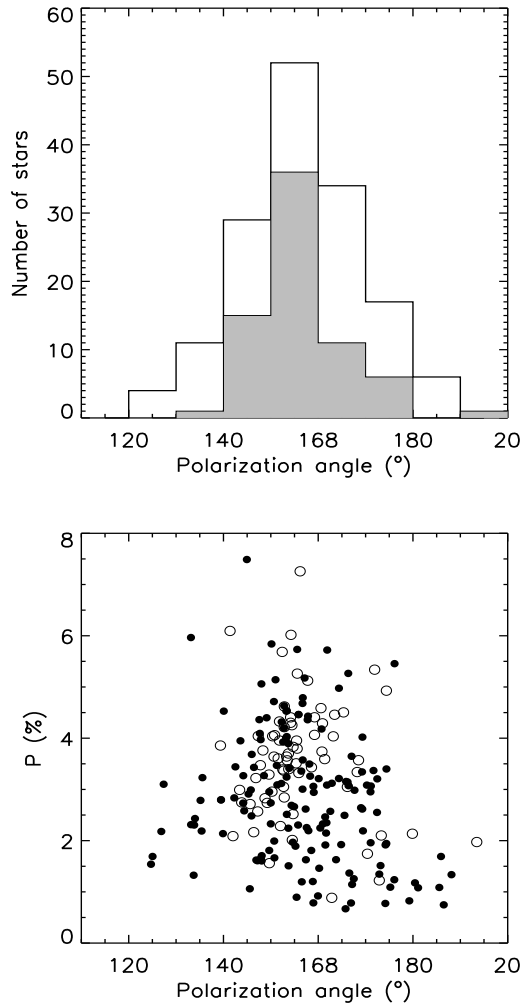}}
\caption{Histograms of $\theta_{P}$ (upper panel) and distribution of P with position angles (lower panel) of the stars corresponding to the two populations towards L1415 dominated by two velocity components with $V_{LSR}$ -5.5$~$km$~s^{-1}$ and -0.5$~$km$~s^{-1}$. The two populations are shown using open and filled histograms in upper panel and open and filled circles in lower panel, respectively.}\label{Fig:two_pop}
\end{figure}

We considered both the components of CO emissions important as they coincide with positions of the cloud where we made the optical polarimetric observations. We compared the polarization values of the stars corresponding to the two populations where two different velocity $^{12}$CO components are dominant. The polarization results of these two populations are consistent. The mean values of the P corresponding to two samples are found to be 2.9$\%$ and 3.4$\%$, respectively and that of $\theta_{P}$ are found to be 157$^{\degree}$ in each sample. The histograms of $\theta_{P}$ and distribution of P with $\theta_{P}$ corresponding to these two populations are shown in Fig. \ref{Fig:two_pop}.

\subsection{L1389} 

\begin{figure}
\resizebox{8cm}{13cm}{\includegraphics{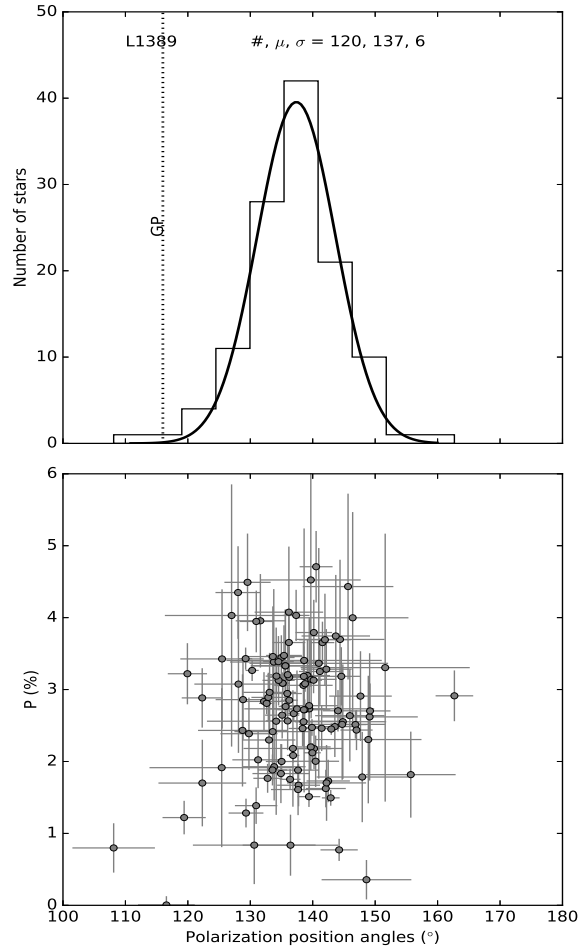}}
\caption{{\bf Upper panel}: Gaussian fitted histogram of $\theta_{P}$ with bin size $10\degree$ in L1389. The orientation of Galactic plane at the latitude of the cloud is shown using dotted line. {\bf Lower panel}: Distribution of P with $\theta_{P}$ of the stars measured towards L1389 shown with filled circles. The submm polarization measurements in L1389 using SCUBA-POL \citep{2009ApJS..182..143M} are also plotted using filled star symbols.} \label{Fig:correctGaussL1389}
\end{figure}

\begin{figure}
\resizebox{8.4cm}{7.9cm}{\includegraphics{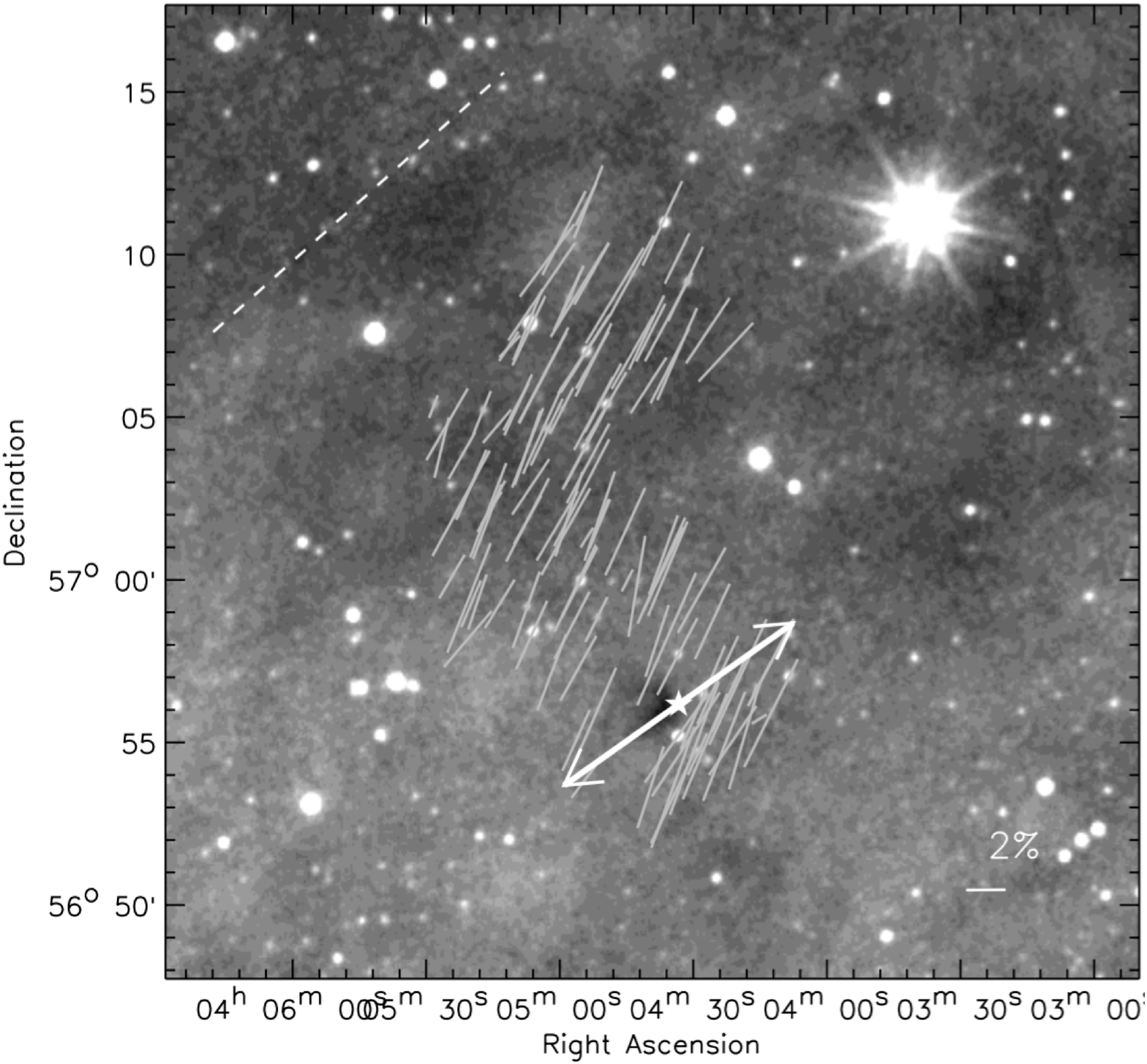}}\\
\resizebox{8.6cm}{7.8cm}{\includegraphics{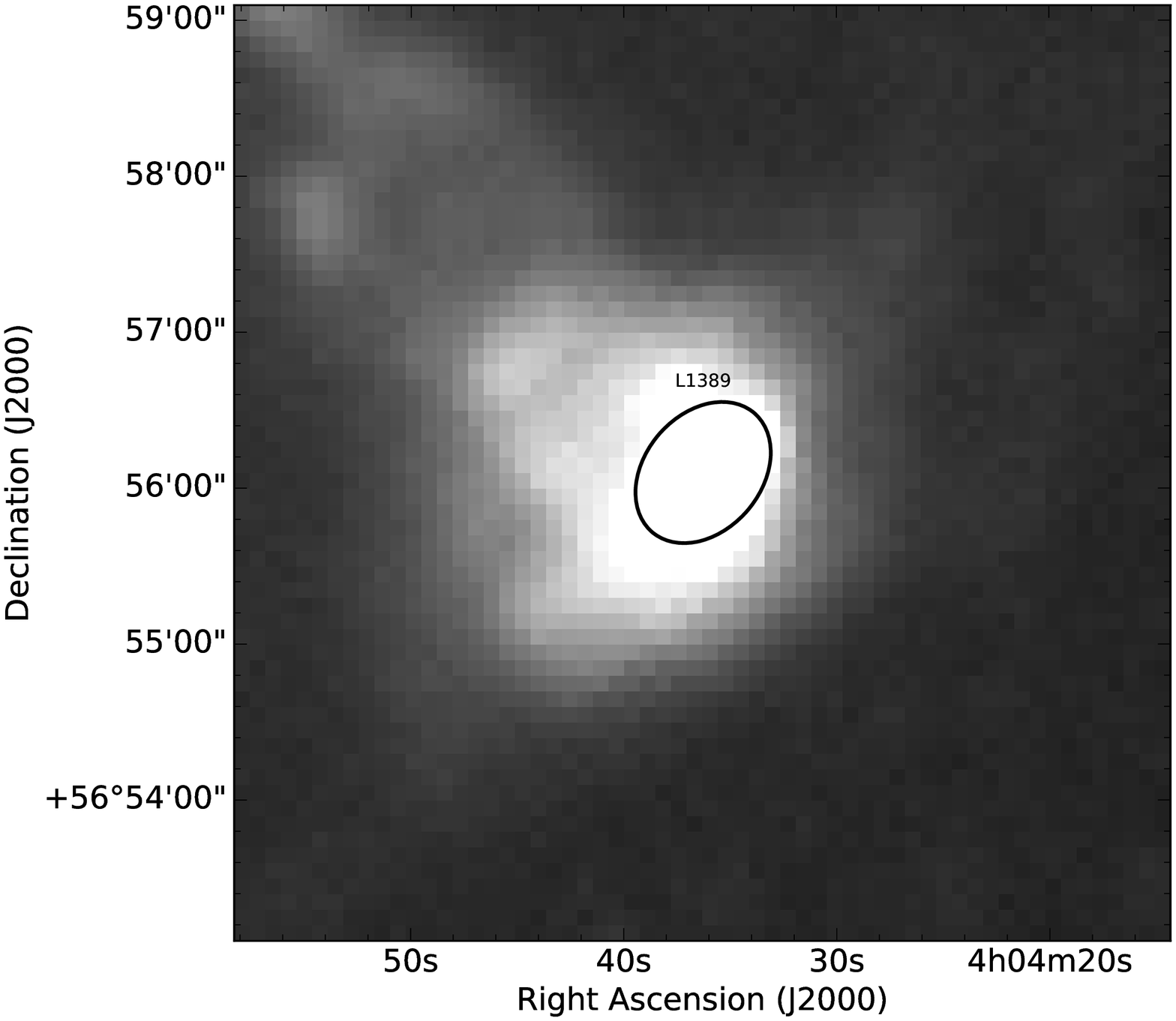}}
%\resizebox{8.5cm}{8.1cm}{\includegraphics{L1389_EPoS.eps}}
\caption{{\bf Upper panel}: Polarization vectors of the stars observed towards L1389 plotted on $0.5^\circ$ $\times$ $0.5^\circ$ WISE 12$~\mu$m images after subtracting the foreground polarization contribution. The double arrow headed line shows the direction of CO outflow from L1389-IRS (shown by star symbol) and the broken line shows the orientation of the Galactic plane. The vector with 2$\%$ polarization is shown as reference. {\bf Lower panel}: Ellipses fitted to the Herschel SPIRE 250$~\mu$m image of L1389 for estimating the minor axis position angle using CSAR.}\label{Fig:correctL1389}
\end{figure}

L1389 is a well studied cloud but lacks the magnetic field information. We present the complete magnetic field morphology of L1389 by carrying out the polarization measurements of 120 stars projected on it. The Gaussian fitted histogram of $\theta_{P}$ measured in L1389 is shown in the upper panel of Fig. \ref{Fig:correctGaussL1389}. The lower panel shows the variation of P with $\theta_{P}$. The upper panel of Fig. \ref{Fig:correctL1389} shows the optical polarization vectors overlaid on $0.5^\circ\times0.5^\circ$ WISE 12$~\mu$m image of L1389.  A vector with 2\% polarization is shown in the figure as a reference vector. The  dashed line shows the orientation of the Galactic plane with a plane-of-the-sky projection angle of 115$^\circ$. The mean values of P and $\theta_{P}$ with corresponding standard deviations are found to be 3.3$\pm$0.9\% and 137$\pm$6$\degree$, respectively.

L1389 has  submillimeter polarization measurements available from SCUBAPOL (SCUBA polarimeter) in the catalog made by \citet{2009ApJS..182..143M}. Although the sample is poor (only two polarization detection are available) with P/$\sigma_{P}$ $\geq$ 2. We have adopted these measurements to assume a mean magnetic field morphology in the high density region of L1389 core. The average values of the P and that of $\theta_{P}$ with their corresponding standard deviation are found to be 19$\pm$5\% and 151$\pm$10$\degree$, respectively. The elongated dust grains aligned with magnetic fields make the thermal continuum emission polarized along their longer axes. Hence to infer the magnetic field geometry information, the submillimeter polarization vectors have to be rotated by 90$\degree$ \citep{1996ASPC...97..325G, 2003ApJ...592..233W}. 

\citet{2012ApJ...751...89C} studied the $^{12}$CO(2-1) outflow in L1389 and found that CO emission is showing a bipolar morphology as seen in low-mass protostellar outflows \citep{2006ApJ...646.1070A, 2007ApJ...659..479J}. The outflow position angle (measured from north increasing towards east) is estimated to be  $\sim125\degree$. \citet{2012ApJ...751...89C} found that near the source L1389-MMS, the redshifted and blueshifted emissions show long and narrow structures ($\sim$8500 AU and $\sim$7500 AU in size, respectively). These structures are found to be extending in the east–west direction and overlapping on each other. \citet{2012ApJ...751...89C} gave various possibilities for this extended emission and considered that the molecular outflow is driven by L1389-MMS, as the most likely scenario. This is different from the structure of bipolar outflows from L1389-IRS. 

\citet{2012ApJ...751...89C} studied the L1389 in submm wavelength using SCUBA and mapped the dense core in 850$~\mu$m emission. Plane-of-the-sky position angle of the major axis of the core measured from 850$~\mu$m emission map is $\sim135\degree$. Hence the minor axis position angle of the L1389 core is $\sim45\degree$. Similar to L1415, we performed the CSAR analysis on the available 250$~\mu$m Herschel-SPIRE image of L1389 also. We considered the head part of this cometary shaped cloud where the IRAS source is detected \citep{2012ApJ...751...89C} for the CSAR analysis. The ellipse is fitted to the round shaped head part where the dense core with L1389-IRS is embedded. The lower panel of Fig. \ref{Fig:correctL1389} shows the ellipse fitted to L1389 Herschel image. The minor axis position angle of the fitted ellipse is found to be $\sim50\degree$ which is similar to the value found in 850 $~\mu$m emission map of L1389 by \citet{2012ApJ...751...89C}. We adopted this position angle for our further analysis of the results in L1389 core.
 
%********************************************************************

%*********************************************************************************
%************************   DISCUSSION   ***************************************** %**********************************************************************************

\section{Discussion}\label{discuss}

 \subsection{Distance of the clouds}\label{sub:dist}
\begin{figure}
\resizebox{8.5cm}{12cm}{\includegraphics{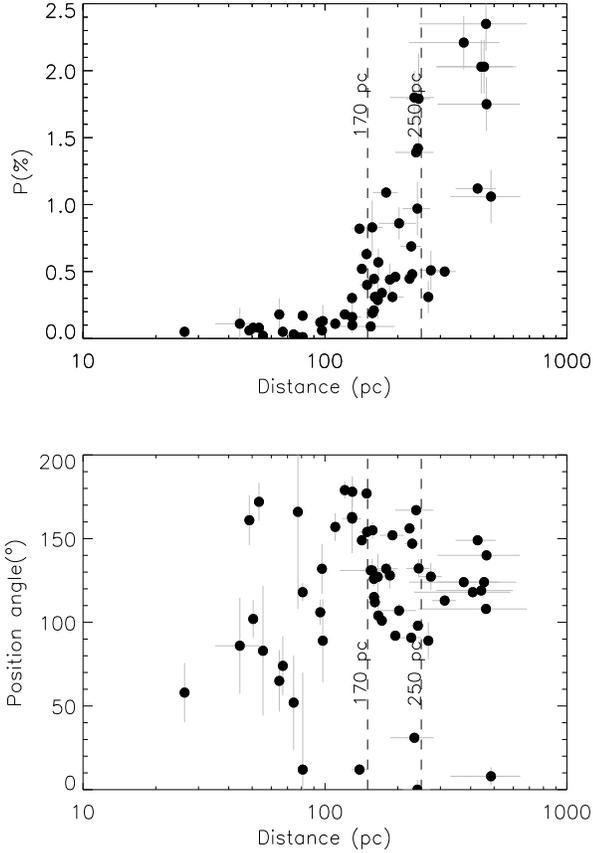}}
\caption{Variation of degree of polarization (upper panel) and position angle (lower panel) of the stars taken from Heiles catalog \citep{2000AJ....119..923H} in a region of $15\degree$ around L1415, with their distances shown. The distance of these stars are measured using the parallax measurements given by \citet{2007A&A...474..653V}.}\label{Fig:heiles_15d}
\end{figure}

 L1415 is located in between a complex of dark clouds L1387 to L1439 which covers $\sim10\degree$ at the boundaries between Camelopardalis and Auriga. \citet{1987ApJ...322..706D}, \citet{1996ApJ...458..561D} and \citet{2003ApJS..144...47B} published the surveys of molecular $^{12}$CO emission in Camelopardalis and the neighboring regions. The molecular clouds in Cam OB1 layer were found with velocities -5 to -20$~$km$~s^{-1}$ \citep{1996ApJ...458..561D}. The Camelopardalis clouds are located almost at the same distance as the Taurus-Auriga star-forming region \citep{Straizys2008}. The CO line survey of \citet{1987ApJ...322..706D} shows that along the line of sight towards L1415 there are two emission components, stronger one at $V_{LSR}$ -5.2$~$km$~s^{-1}$ and weaker one at 0 km$~s^{-1}$. In our TRAO observations of L1415 using $^{12}$CO line, we detected these two components at $V_{LSR}$ -5.5$~$km$~s^{-1}$ and -0.5$~$km$~s^{-1}$, respectively. Hence L1415 can reasonably be assumed as the part of Camelopardalis complex.  An upper limit of the distance of L1415 can be assumed based on to stellar population model of \citet{2003A&A...409..523R}. This model predicts the number of foreground stars based on to the the limiting magnitude and distance dependence of the cumulative number of stars in the cloud direction finding a maximum distance of 190 pc for an average value of 0.5 foreground stars and 250 pc for 1 foreground star. The results were consistent with the study of L1415 by \citet{1981ApJS...45..121S}. In this work we tried to find the distance of L1415 with the help of stars projected towards the cloud for which polarization measurements are available in the Heiles catalog \citep{2000AJ....119..923H}. We selected the stars from this catalog lying within a region of 15$^{\degree}$ around L1415. Fig. \ref{Fig:heiles_15d} shows the variation of the degree of polarization and polarization position angles of these stars from Heiles catalog \citep{2000AJ....119..923H} with their distances. The distance of these stars are calculated using the parallax measurements given by \citet{2007A&A...474..653V}. A sudden change in the polarization values of these stars can be noticed within the distances 170 pc and 250 pc suggesting the distance of the cloud is somewhere in this range. We have adopted the distance of L1415 as 250 pc. The analysis supporting this distance of L1415 is explained in section \ref{subsec:fg_sub}.

The distance of L1389 is reported ranging from 210 pc \citep{2007A&A...474..653V} to 300 pc \citep{1987ApJ...322..706D}. The distance was derived by \citet{1997A&A...326..329L} based on association in projected space and radial velocities with other clouds in Lindblad Ring. The Lindblad Ring structures have a mean distance of $\sim$300 pc \citep{1987ApJ...322..706D} in the direction of L1389. A bright star HD$~$25347 with spectral type G5 III at a distance of 210$\pm$40 pc \citep{2007A&A...474..653V}, is located approx. 11$\arcmin$ (0.65 pc at 200 pc) to the south of L1389 cloud. This star could be responsible for the cometary shape of the cloud and cloudshine from the rim if it is located at similar distance to that of the globule. Assuming the possible association with HD25347 and Lindblad ring, \citet{2010ApJS..188..139L} estimated a distance of 250$\pm$50 pc for L1389. We have also adopted the distance of L1389 as $\sim$250 pc for our further analysis.

\subsection{Subtraction of interstellar polarization}\label{subsec:fg_sub}

The polarization measurements provide magnetic field morphology in the plane-of-the-sky averaged over total line-of-sight and weighted by the density and the alignment efficiency of the dust grains. But, in principle, to obtain the actual magnetic field morphology in a cloud, the foreground polarization contribution has to be excluded. From the previous polarization studies by us towards the clouds located at distances less than $\sim$500 pc \citep[e.g., ][]{2013MNRAS.432.1502S, 2016A&A...588A..45N} and from the literature \citep{2009ApJ...704..891L}, it was found that the change occurred to the values of the $\theta_{P}$ after correcting for the foreground contribution is not noteworthy specially if the observed P is relatively higher (i.e. $\gtrsim 1\%$). To test this fact more clearly, we subtracted the foreground contribution from L1415 (at a distance of $\sim$250 pc) polarization measurements in this work. 

We made a search of stars that are located within a circular region of $2.0\degree$ radius around L1415 and with their parallax measurements available from Hipparcos satellite \citep{2007A&A...474..653V}. We rejected the peculiar stars (emission line stars, stars in a binary or multiple system or are peculiar according to the information provided by the Simbad) and considered the normal stars for subtracting the polarization component due to the foreground material from our observed values. Eleven stars are found for which the parallax measurements are available in the catalog produced by \citet{2007A&A...474..653V}. We selected only those stars for which error in parallax and the parallax values are $\leq$ 0.5. These selected eleven stars are shown in Table \ref{tab:fg}. Similar to the target stars, these stars were also observed in R-band using AIMPOL.

\begin{figure}
\resizebox{9cm}{12cm}{\includegraphics{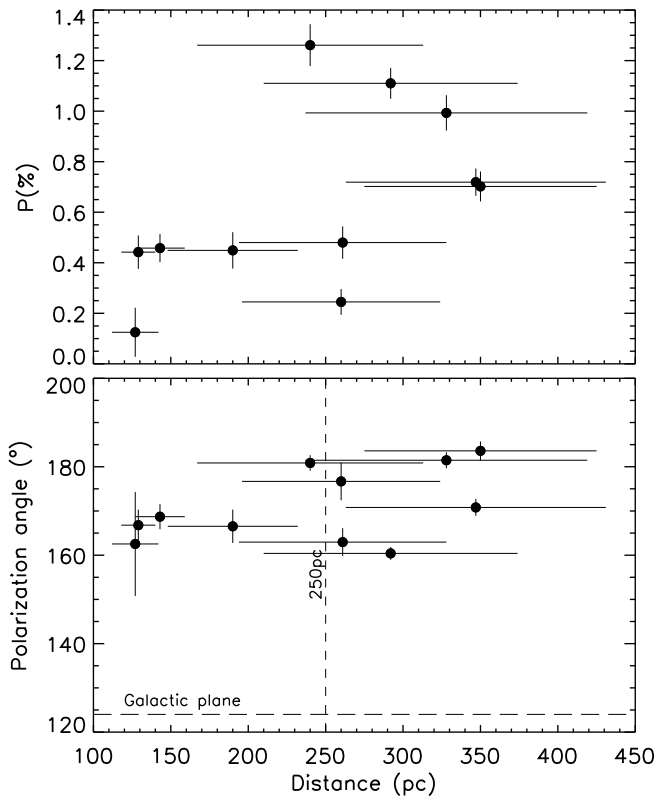}}
\caption{Variation of degree of polarization and position angle of foreground stars observed toward L1415 with their distances. The distances of these stars are estimated using the parallex measurements gven by \citet{2007A&A...474..653V}.}\label{Fig:PvsDist}
\end{figure}

The polarimetric results of the foreground stars are shown in Table \ref{tab:fg}. Distribution of the degree of polarization and position angle of these stars is shown in the lower panel of Fig. \ref{Fig:correctGaussL1415} shown with filled star symbols. The variation of P and $\theta_{P}$ of these stars with their corresponding distances are given in Fig. \ref{Fig:PvsDist}. The distances to the eleven stars foreground to L1415 range from $\sim127$ pc to $\sim350$ pc. As expected, the degree of polarization is found to increase with the distance. There is a jump in degree of polarization at a distance of at around 240 pc. Since the distance of L1415 is assumed to be 250 pc, therefore we considered the four stars towards L1415 which are at distances less than the 240 pc for the foreground polarization subtraction from target stars. The star at 240 pc and 1.26\% polarization falls in the uncertainty limits of the distance of L1415 hence we did not consider this star for subtraction.

%********************************************************************************************    
%\input{table3_FG}    % details of FG stars
\begin{table}
\caption{$R_{c}$-band polazisation results of foreground stars towards L1415 and L1389.}\label{tab:fg}
\begin{tabular}{llllll}\hline
Id&Star	Name	&V   	& P $\pm$ $\sigma_P$ & $\theta$ $\pm$ $\sigma_{\theta}$  &D$^{\dagger}$\\ 
  &				&(mag)	& (\%) 			&($\degree$)	&(pc) \\  \hline
1 &HD$~$30136	           &6.8 &0.12$\pm$0.09&  163$\pm$11 & 127\\
2 &HD$~$30696	           &7.9 &0.44$\pm$0.06&  167$\pm$3  & 129\\
3 &HD$~$30583	           &7.7 &0.45$\pm$0.05&  169$\pm$2  & 143\\
4 &HD$~$30480	           &8.9 &0.44$\pm$0.07&  167$\pm$3  & 190\\
5 &BD$+$54$\degree$792 &9.9 &1.26$\pm$0.08&  181$\pm$1  & 240\\
6 &HD$~$232997	       &9.0 &0.24$\pm$0.05&  177$\pm$4  & 260\\
7 &HD$~$29945	           &8.3 &0.48$\pm$0.06&  163$\pm$3  & 261\\
8 &HD$~$30326	           &8.4 &1.11$\pm$0.06&  160$\pm$1  & 292\\
9 &HD$~$29720	           &8.5 &0.99$\pm$0.07&  181$\pm$1  & 328\\
10 &HD$~$21846	       &8.2 &0.71$\pm$0.05&  171$\pm$1  & 347\\
11 &HD$~$22181	       &7.3 &0.70$\pm$0.05&  184$\pm$2  & 350\\
\hline
\multicolumn{6}{c}{Polarisation results for 2 foreground stars towards L1389}\\
\hline
1 &HD$~$25641	           &6.8 &0.20$\pm$0.10&  163$\pm$11 & 75\\
2 &HD$~$25021	           &8.9 &0.16$\pm$0.07&  167$\pm$3  & 140\\
\hline
\end{tabular}

$^{\dagger}$ distances are estimated using the Hipparcos parallax measurements taken from \citet{2007A&A...474..653V}.
\end{table}
%********************************************************************************************    
We calculated the mean value of the Stokes parameters corresponding to these stars i.e., $Q_{fg}$ (=Pcos2$\theta$) and $U_{fg}$ (=Psin2$\theta$) using the observed values of the degree of polarization and the position angles. The Stokes parameters thus estimated are found to be  $Q_{fg}$=0.330 and $U_{fg}$=$-$0.162. For removing the foreground contribution of interstellar polarization from our observed values, the Stokes parameters corresponding to the target stars, $Q_{\star}$ and $U_{\star}$ were calculated. The intrinsic polarization of target stars are represented by Stokes parameters $Q_{i}$ and $U_{i}$ and calculated using the relations 
\begin{equation} \label{qu_star_ism}
Q_{i}=Q_{\star} - Q_{fg},\\
U_{i}=U_{\star} - U_{fg}
\end{equation} 
We then estimated the intrinsic degree of polarization $P_{i}$ and position angle $\theta_{i}$ of the target stars using the equations
\begin{equation} \label{ppa_star_ism}
P_{i}=\sqrt{(Q_{i})^2+(U_{i})^2},\\
\theta_{i}=0.5\times tan^{-1}\left(\frac{U_{i}}{Q_{i}}\right)
\end{equation}
We did not notice a significance change in the polarization values after correcting for the foreground contribution. This may be because the mean value of P in foreground stars is very small (i.e. 0.63\%). We observed two stars foreground to L1389 (see Table \ref{tab:fg}) and followed the similar procedure for the foreground polarization subtraction from L1389. We did not notice any significant change in the polarization measurements of L1389.

%############################FG starswith distances#############################

\subsection{Outflow direction and magnetic fields}

The angular offsets between the various quantities in five cores with Very Low Luminosity Objects \citep[VeLLOs;][]{2015A&A...573A..34S} and the two cores studied in this work are shown in Table \ref{tab:offsets}. This table shows the the angular offset between core scale magnetic field ($\theta^{sub}_{B}$, obtained from the sub-mm polarization measurements), envelope magnetic field ($\theta^{opt}_{B}$,  obtained from the optical polarization measurements), outflow directions ($\theta_{out}$) and minor axis ($\theta_{min}$) of the cloud cores. The caution must be taken while interpreting these angular offsets as calculation are solely made between two projected quantities. 

Both the clouds L1415 and L1389 have been found to be associated with collimated outflow activities from the central sources. The embedded L1415-IRS is found to be associated with HHOs \citep{2007A&A...463..621S}. HHOs  \citep{1950ApJ...111...11H, 1952ApJ...115..572H} are the tracers of pre-main-sequence stars \citep{2001ARA&A..39..403R}. The presence of HHOs in the deeply embedded young stellar objects makes the association of jets evident and the radial velocity similarities of HHOs suggest that the outflow orientation is almost close to the plane-of-the-sky \citep{2007A&A...463..621S}. The NIR morphology of L1415-IRS is indicating a bipolar outflow seen in 2.16$~\mu$m \citep{2007A&A...463..621S}. If we join the positions of HHOs and the L1415-IRS with a line,  the outflow axis in the plane-of-the-sky is found to be with position angle as $\sim10\degree$ (from north increasing towards east).  In L1389, the position angle of outflow is measured using $^{12}$CO(2-1) molecular line observations \citep{2012ApJ...751...89C}. The outflow in L1389 is found to be well collimated with a plane-of-the-sky position angle of $\sim125\degree$ \citep{2012ApJ...751...89C}. The position angle of outflows measured in the plane-of-the-sky can be uncertain by $\sim10-15\degree$ as noticed in our previous study \citep{2015A&A...573A..34S}.

\begin{figure}
\resizebox{9.0cm}{7.0cm}{\includegraphics{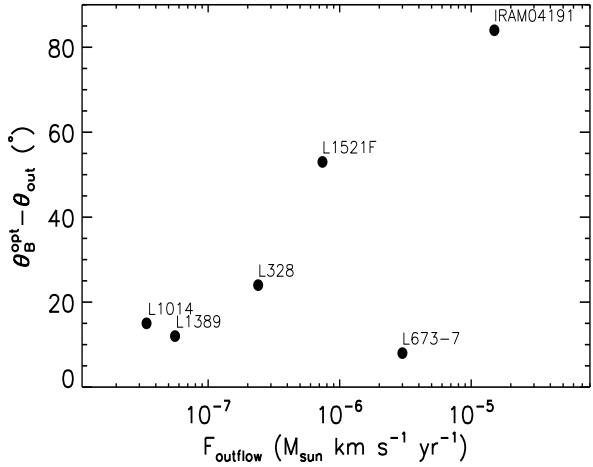}}
\caption{Variation of offset between $\theta^{opt}_{B}$ and $\theta_{out}$ in the cores with VeLLOs \citep{2015A&A...573A..34S} and the L1389 (this work) with the outflow momentum flux. The momentum flux values of these sources are adopted from literature i.e. IRAM04191; \citep{1999ApJ...513L..57A}, L1521F; \citep{2013ApJ...774...20T}, L328; \citep{2013ApJ...777...50L}, L673-7; \citep{2012AJ....144..115S}, L1014; \citep{2005ApJ...633L.129B} and L1389; \citep{2012ApJ...751...89C}.}\label{Fig:correlation}
\end{figure}

The angular offsets between $\theta^{opt}_{B}$ and $\theta_{out}$ for L1415 and L1389 are 35$\degree$ and 12$\degree$, respectively (column 7 of Table \ref{tab:offsets}). The mean value of these offsets is $23\degree$. The previous study of magnetic field in four cores with VeLLOs \citep{2015A&A...573A..34S} reported the mean offset of $25\degree$. The mean value of offset from previous study and this study is $24\degree$. The outflows from protostars are thought to generate turbulence in the surrounding medium making the relatively weak magnetic fields scrambled which can further misalign the core and envelope magnetic fields. The estimated outflow parameters for low luminosity objects suggest that these objects have outflows as highly compact, of least mass and with least energy as compared to the outflows from already known Class 0/I sources \citep[e.g.][]{2002A&A...393..927B, 2004A&A...426..503W, 2006ApJ...649L..37B, 2011ApJ...743..201P}. To test this scenario, we checked the distribution of outflow force and the offsets found in the cores with low luminosity objects (see Fig. \ref{Fig:correlation}). The momentum flux values of these sources are adopted from literature i.e. IRAM04191; \citep{1999ApJ...513L..57A}, L1521F; \citep{2013ApJ...774...20T}, L328; \citep{2013ApJ...777...50L}, L673-7; \citep{2012AJ....144..115S}, L1014; \citep{2005ApJ...633L.129B} and L1389; \citep{2012ApJ...751...89C}. The sources with least energetic outflows are expected to have the better alignment between $\theta^{opt}_{B}$ and $\theta_{out}$. However, it should be noted that the values of outflow force can be largely uncertain by many unknown parameters. For instance, \citet{2013ApJ...777...50L} calculated the momentum flux of L328-IRS by assuming an averaged inclination angle (i.e. $\textit{i}$= 57.3$^{\degree}$). In Fig. \ref{Fig:correlation}, if we exclude the case of IRAM0419 \citep[with a possible different environment;][]{2015A&A...573A..34S}, the envelope magnetic fields and the outflows are found to be aligned in L1014, L1389 and L328 cores with relatively lower outflow forces. 

The dependence of alignment of the outflow, with local magnetic field at $\sim10$ AU scale, on the magnetic field strength is tested by \citet{2004ApJ...616..266M}  based on the MHD simulations. Their study suggest that in a slowly rotating, magnetized molecular cloud core undergoing gravitational collapse, the alignment is independent of the magnetic field strengths assumed in simulations. But the dependence of alignment on the magnetic field strength is considered on the cloud scale. A better alignment between outflows and magnetic field orientation is noticed in stronger magnetic fields. But such correlation is not noticed in the studies made to test the relative alignment between outflows and envelope magnetic fields such as in T-Tauri stars in the Taurus molecular cloud \citep{2004A&A...425..973M}. \citet{2011ApJ...743...54T} used the optical polarimetric technique and found the randomly distributed offsets between magnetic fields and outflow orientations in the protostars with different ages. However \citet{2013ApJ...770..151C} noticed a statistical evidence of alignment in the study of class 0/I sources. The lack of alignment between outflows and envelope magnetic fields in relatively evolved T-Tauri stars may be present due to the injection of more turbulence from the outflows to their surrounding resulting into randomization of magnetic fields \citep{2013ApJ...770..151C}. The low luminosity sources are expected to inject the least possible turbulence into their surroundings and hence causing lowest scrambling of the magnetic fields. This may result into clouds retaining their initial field morphologies. In this work and our previous work on VeLLOs, we found an alignment in the outflows and magnetic fields.

\subsection{Magnetic field strength using classical CF technique}

We attempted to understand the relation between magnetic field strength and the alignment of outflows and magnetic fields towards the clouds with low luminosity stars. We used the updated Chandrasekhar-Fermi (CF) relation ($B_{pos}=9.3\sqrt{n(H_{2})}\delta v/\delta\theta$) \citep{1953ApJ...118..113C, 2001ApJ...546..980O, 2005mpge.conf..103C} to estimate the plane-of-the-sky component ($B_{pos}$) of the magnetic field strengths in L1415 and L1389. In this relation $n(H_{2})$ shows the number density of the molecular hydrogen which is the major constituent of the clouds, $\delta v$ represents the velocity dispersion and the dispersion in $\theta_{P}$ is shown using $\delta\theta$. From our TRAO observations, we checked the line width values at various positions (coinciding with our polarization observation positions) of $^{12}$CO(1-0) emission in L1415 and found that these values vary only by $\Delta v$ = 0.05$~$km$~s^{-1}$. Hence we measured the average $^{12}$CO(1-0) line widths, $\Delta v$ = 1.65$\pm$0.03$~$km$~s^{-1}$ and 1.64$\pm$0.02$~$km$~s^{-1}$ corresponding to the two components in line of sight towards L1415, respectively.  We estimated the column density ($N(H_{2})$) from the extinction using the relation $N(H_{2})/A_{V}= 9.4\times 10^{20} cm^{-2} mag^{-1}$ from \citet{1978ApJ...224..132B}. The average value of extinction traced by the stars lying behind the cloud (distance $\geq$250 pc) observed in this study is found to be $\sim$0.9 mag. The angular diameter of the cloud is found to be $\sim 22\arcmin$. Considering 250 pc as the distance to L1415, the value of $n(H_{2})$ is calculated to be $\sim 200~cm^{-3}$. We used this value of volume density to estimate the magnetic field strength in the regions upto which optical observations were made in L1415. The $\delta\theta$ used to estimate the field strength is calculated from the standard deviation in $\theta_{P}$ which was obtained by Gaussian fitting to the position angles. We have corrected the dispersion in $\theta_{P}$ by uncertainty in $\theta_{P}$ \citep{2001ApJ...561..864L, 2010ApJ...723..146F}. We adopted the procedure explained by \citet{2010ApJ...723..146F} according to which the dispersion in position angles is corrected in quadrature by the polarization angle using $\Delta\theta = ({\sigma_{std}}^{2} - {\langle\sigma_{\theta}\rangle}^{2})^{1/2}$, where the mean error $\langle\sigma_{\theta}\rangle$ was estimated from $\langle\sigma_{\theta}\rangle = \Sigma\sigma_{\theta i}/N$, where $\sigma_{\theta i}$ is the estimated uncertainty in the star's polarization angle.\footnote{The uncertainty in the position angles is calculated by error propagation in the expression of polarization angle $\theta$, which gives, $\sigma_{\theta} = 0.5\times\sigma_{P}/P$ in radians, or $\sigma_{\theta} = 28.65\degree\times\sigma_{P}/P$ \citep[see;] []{Serkowski1974} in degrees.} Considering these values, we estimated a magnetic field strength in the regions of L1415 where two components of CO emission with different $V_{LSR}$ are detected. The magnetic field strengths in these regions are found to be $\sim30\pm18~\mu$G and $\sim25\pm11~\mu$G, respectively. The uncertainties in the magnetic field strengths have been measured using the uncertainties in velocity dispersion and the position angles. Thus the mean value of the magnetic field strength in L1415 is $\sim28\pm15~\mu$G. The CO line width information for L1389 ($\Delta v$ = 2.0$~$ km$~s^{-1}$) is adopted from \citet{2012ApJ...751...89C}. Using the similar estimation method and using the polarization data on head part of L1389, we obtained a magnetic field strength of $\sim149~\mu$G in L1389 (uncertainty in the $B_{pos}$ is not estimated here because the uncertainty in the velocity dispersion is not known). The typical uncertainty ($\sigma B_{pos}$) in the field strength is found to be $\sim0.5B_{pos}$ by considering the uncertainties in $\theta_{P}$ and dispersion velocity. Based on to the present study and the study of magnetic field in cores with VeLLOs by \citet{2015A&A...573A..34S}, we noticed the alignment is better in the cores with stronger magnetic fields but our finding is not statistically significant.

\subsection{Caveats of CF technique}
The CF technique has been used as a convenient tool to estimate the ISM magnetic field strength but the results estimated from this technique carries significant uncertainties. The basic assumption of the CF method is that it is applicable in the cases where $\delta\theta~<~25\degree$ \citep{2001ApJ...546..980O}. This assumption is the limitation in using this classical method. \citet{2001ApJ...546..980O} and \citet{ 2001ApJ...561..800H} for the first time show that this technique results into overestimation of the field strength over a coarser resolution. According to recent consensus, structure function (SF) analysis is proved to be a powerful statistical tool to estimate the magnetic field intensity on finer resolutions and relation between large scale and turbulent component of the magnetic fields in the molecular clouds \citep{2008ApJ...679..537F, 2009ApJ...696..567H, 2009ApJ...706.1504H, 2010ApJ...723..146F}. This method suggests that the analysis of small scale randomness in the magnetic field lines could estimate the magnetic field strength.

%--------------------SF plot----------------------------------------------
\begin{figure}
\resizebox{8.0cm}{7.0cm}{\includegraphics{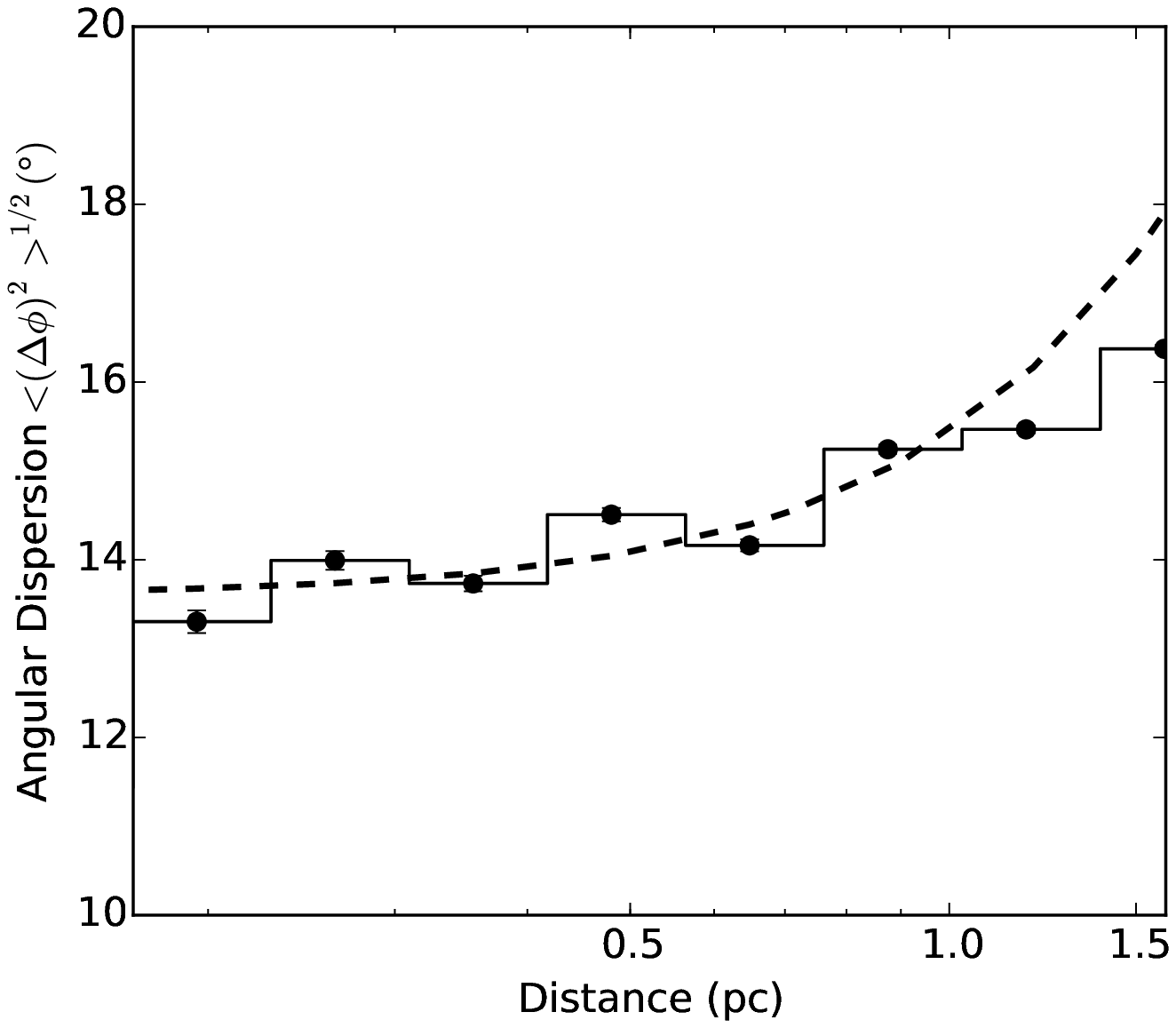}}\\
\resizebox{8.0cm}{7.0cm}{\includegraphics{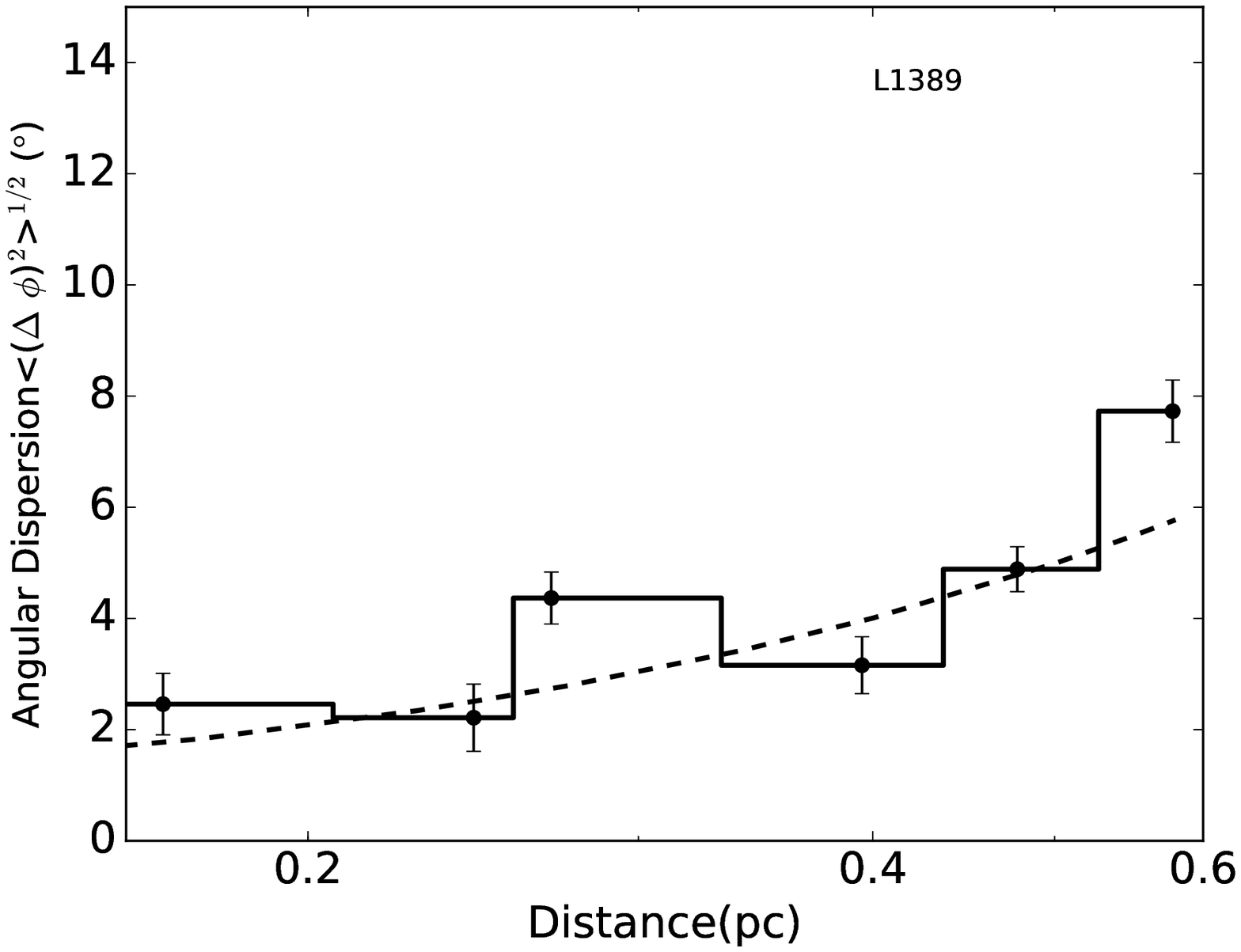}}
%\resizebox{8.5cm}{8.1cm}{\includegraphics{L1389_EPoS.eps}}
\caption{Distribution of angular dispersion function of the polarizations angles ($\langle\Delta(\phi)^{2}\rangle^{1/2} (\degree)$) with distance for 224 stars observed towards L1415 (upper panle) and 120 stars in L1389 (lower panel). The dashed line shows the best fits to the data upto 1.5 pc in L1415 and 0.4 pc in L1389, respectively.}\label{Fig:SF}
\end{figure}

%--------------------SF plot----------------------------------------------

\subsection{Magnetic field strength using structure function analysis}
Structure function of the polarization maps can be used as a technique to measure the magnetic field intensity and to probe the small scale perturbations using turbulent angular dispersion \citep{2008ApJ...679..537F, 2009ApJ...696..567H}. It infers the behavior of the dispersion of the polarization angles as a function of the length scale in molecular clouds. The net magnetic field, $B_{o}(x)$ basically consists of large scale regular magnetic fields and turbulent component, $B_{t}(x)$.  The SF is a second order quantity which is a function of polarization angle and is defined as the average of the squared difference between the polarization angle measured at two positions separated by \textit{l} \citep{2008ApJ...679..537F}  and the square root of this quantity is known as angular dispersion functions (ADF) and measured as shown in eq. \ref{adf}:

\begin{equation} \label{adf}
\langle\Delta(\phi)^{2}(\textit{l})\rangle^{1/2} = \Bigg\{ \frac{1}{N(\textit{l})} \sum_{n=i}^{N(\textit{l})}[\phi(x) - \phi(x+\textit{l})]^{2} \Bigg\}^{1/2}
\end{equation}

The SF analysis can lead us to understand the magnetic field intensities at various scales and hence helps in understanding the small scale turbulence. In other words SF can possibly be used as a tool to correlate the small scale turbulent eddies to large scale magnetic fields \citep{2008ApJ...679..537F, 2009ApJ...696..567H, 2010ApJ...723..146F, 2012ApJ...747...21S, 2016A&A...588A..45N}. The flat profile of the SF suggests the size of the smallest turbulent cells \citep{2008ApJ...679..537F}.

The structure function in range $\delta < \textit{l} << d$ (where $\delta$ and \textit{d} are correlation lengths which characterizes $B_{t}(x)$ and $B_{o}(x)$, respectively) can be estimated with expression shown in eq. \ref{adf1}.

\begin{equation}\label{adf1}
\langle\Delta(\phi)^{2}(\textit{l})\rangle_{tot}  \simeq \textit{b}^{2} + \textit{m}^{2}\textit{l}^{2} + \sigma_{M}^{2}(\textit{l})
\end{equation}

In eq. \ref{adf1}, $\langle\Delta(\phi)^{2}(\textit{l})\rangle_{tot}$ represents the total measured dispersion from the data. Here $\sigma_{M}^{2}(\textit{l})$ are the measurements uncertainties as a function of \textit{l} and estimated by taking the mean of the variances on $\Delta\phi(\textit{l})$ in each bin. In the equation, quantity $\textit{b}^{2}$ is the constant turbulent contribution measured by the intercept of the fit to the data after subtracting $\sigma_{M}^{2}(\textit{l})$. The quantity $\textit{m}^{2}\textit{l}^{2}$ is a smoothly increasing contribution along with length $\textit{l}$ where $\textit{m}$ represents the slope of this linear pattern. The ratio of turbulent component and large scale magnetic fields can be estimated as:

\begin{equation}\label{ratio}
\frac{\langle B_{t}^{2}\rangle^{1/2}}{B_{o}} = \frac{\textit{b}}{\sqrt{2-\textit{b}^2}}
\end{equation}

For the clouds studied in the work, we estimated ADF and plotted it as a function of distance as shown in Fig. \ref{Fig:SF}. The uncertainties though very less are also potted. In the figure, the bins represent the $\sqrt{\langle\Delta(\phi)^{2}(\textit{l})\rangle_{tot} - \sigma_{M}^{2}(\textit{l})}$ which is ADF corrected by uncertainty. In both the clouds, we estimated net turbulent component to the angular dispersion, $\textit{b}$ and the ratio of turbulent component and the large scale magnetic fields using eq. \ref{ratio}. The strength of the plane-of-the-sky component of magnetic fields ($B_{pos,modCF}$) is estimated from modified CF relation \citep{2015ApJ...807....5F} using eq. \ref{Bmod}. We performed the SF technique on the polarization data towards head part of L1389 because the core on the head part contains the embedded source and the velocity dispersion values are measured towards the head part only \citep{2012ApJ...751...89C}. The turbulent contribution to the angular dispersion, $\textit{b}$ is found to be 13.6$\pm$0.3$\degree$ (0.24$\pm$0.01 rad) and 1.9$\pm$0.4$\degree$ (0.03$\pm$0.007 rad) in L1415 and L1389, respectively. The values of ratio of the turbulent component and the large-scale magnetic fields estimated using eq. \ref{ratio} in L1415 and L1389 are found to be 0.14$\pm$0.007 and 0.02$\pm$0.004, respectively. The estimated values towards the two clouds studied here are shown in Table in \ref{tab:CFSF}.

\begin{equation}\label{Bmod}
B_{pos, modCF} = 9.3\Bigg[\frac{2~n_{H_{2}}}{cm\textsuperscript{-3}}\Bigg]^{1/2} \Bigg[\frac{\Delta V}{km~s\textsuperscript{-1}}\Bigg] \Bigg[\frac{\textit{b}}{1^{\circ}} \Bigg]^{-1} \mu G
\end{equation}

The magnetic field strength estimated in L1415 and L1389 are found to be consistent in both CF and SF techniques. The values obtained for $B_{pos, modCF}$ should be considered as the rough estimation because of the large uncertainties involved in quantities used in its measurement. Along with the uncertainties in the estimation of velocity dispersion and $\textit{b}$, the large errors present in the estimation of the volume density may be the dominant source of uncertainties in plane-of-the-sky magnetic field strength.

%%%%%%%%%%%%%%%%%%%%%%%
%\input{table_CFSF}

\begin{table}
\begin{small}
  \caption{Parameters estimated from SF analysis.}\label{tab:CFSF}
  \begin{tabular}{llll}
\hline
Cloud          & $\textit{b}$ ($^{\circ}$) & $\frac{\langle B_{t}^{2}\rangle^{1/2}}{B_{o}}$ & $B_{pos, modCF} (\mu G)$\\
\hline
 L1415              & 13.6  & 0.17 & 23 \\  
 L1389               & 1.9  & 0.02 & 140 \\  
\hline
\end{tabular}
\end{small}
\end{table}
%%%%%%%%%%%%%%%%%%%%%%%

\subsection{Statistical view on the magnetic fields of cores with low luminosity protostars}

%******************************
%\input{table4_stats}
\begin{table*}
\centering
\begin{minipage}{\textwidth}
\caption{Mean magnetic field position angles and angular offsets between magnetic fields, cloud minor axis, and outflows in the core with low luminosity objects. We have adopted some of the quanties (shown with $^{\dagger}$) from our previous work in \citet{2015A&A...573A..34S} for the purpose of increasing our sample in the present study.}\label{tab:offsets}
\scriptsize
\begin{tabular}{llccccccccc}
\hline
Cloud Id.

& $\theta^{opt}_{B}$
& $\theta^{sub}_{B}$		
& $|\theta^{opt}_{B}-\theta^{sub}_{B}|$
& $|\theta^{opt}_{B}-\theta_{min}|$
& $|\theta^{sub}_{B}-\theta_{min}|$
& $|\theta^{opt}_{B}-\theta_{out}|$
& $|\theta^{sub}_{B}-\theta_{out}|$
& $|\theta_{out}-\theta_{min}|$
& $\sigma_{\theta^{opt}_{B}}$
\\

&(deg)	
&($^{\circ}$)	
&($^{\circ}$)
&($^{\circ}$)	
&($^{\circ}$)	
&($^{\circ}$)	
&($^{\circ}$)
&($^{\circ}$)\\

(1)&(2)&(3)&(4)&(5)&(6)&(7)&(8)&(9)&(10)\\
\hline
 $^{\dagger}$IRAM 04191 & 112 	&44	 	&68		&82  	& 14 	& 84 			& 16  		&2 &  13   \\
 $^{\dagger}$L1521F     & 22  	&--	 	&--		&60		& -- 	& 53 			& -- 		&7 & 13  \\
 $^{\dagger}$L328       & 44  	&--	 	&--		&--		& -- 	& 24 			& -- 		&-- & 21  \\
 $^{\dagger}$L673-7     & 47  	&--	 	&--		&47		& -- 	& 8  			& -- 		&55 & 48 \\
 $^{\dagger}$L1014      & 15  	&--	 	&--		&55		& -- 	& 15 			& --  		&40 & 13  \\
 L1415      & 155   &--     &--     &50    &--     & 35            &--         &15 & 10 \\
 L1389      & 137   &151     &14     &87    &101     & 12            &26         &75 & 06 \\
\hline
\end{tabular}

$opt:$ Optical data; $sub:$ Submm data; $min:$ Minor axis of the clouds; $out:$ Outflow direction from the VeLLO; $\dagger$: \citet{2015A&A...573A..34S}.
\end{minipage}
\end{table*}
%******************************

Among the two cores studied in this work, L1389 has inner magnetic field information inferred from submm polarization observations. The inner and outer magnetic field directions have a offset of $14\degree$ implying that the field lines are anchored in high density region and continues to low density parts of the cloud. The angular offsets between ${\theta_{B}}^{opt}$ and $\theta_{minor}$ in L1415 and L1389 are found to be $50\degree$ and $87\degree$, respectively. The mean value of the offsets between minor axes and magnetic fields in the two cores studied in this work and the four cores with minor axes information from \citet{2015A&A...573A..34S} (Table \ref{tab:offsets}), is found to be $63\degree$. Thus the minor axes and envelope magnetic fields are found to be misaligned in these cores. The magnetically dominated star formation models assume a rather aligned fields and minor axes making the above finding inconsistent to it. \citet{2000ApJ...540L.103B} considered the clouds as triaxial bodies and the idea of random viewing angles. This combination of assumptions can result into the average offset between magnetic fields and minor axes typically falls in the range of 10$\degree$- 30$\degree$. The value of offset as $\sim60\degree$ between ${\theta_{B}}^{opt}$ and $\theta_{minor}$, therefore can not be solely explained based on projection effect.

The offset between $\theta_{out}$ and $\theta_{minor}$ of L1415 and L1389 are found to be $15\degree$ and $75\degree$, respectively. The mean value of the offsets from our previous study of VeLLOs and this study becomes  $32\degree$. Four low-mass cores namely L483, L1157, L1448-IRS2, and Serp-FIR1 with Class 0 protostars were studied by 
\citet{2013ApJ...770..151C} using 350$~\mu$m polarization observations. These observations were done to test the magnetically regulated core-collapse models. A good correlation between outflow direction and the minor axes was found in that study. Such studies allow us to use the inclination angle of the outflow as proxy of the minor axis position angle. 

It has already been discussed by \citet{2015A&A...573A..34S} that magnetic fields are found to be preferabelly aligned with the Galactic plane orientation with a mean offset value of $34\degree$ in the five cores with VeLLOs. In this work also, we noticed that the magnetic fields in L1415 and L1389 are found to be parallel to the Galactic plane (with the mean offset between Galactic plane orientation and magnetic field as $27\degree$). Thus the mean value of the offset between Galactic plane orientation and magnetic field in the seven cores (five studied by us in \citet{2015A&A...573A..34S} and two in this work) is found to be $30\degree$. Many clouds are found coupled to the Galactic plane \citep{1990AJ.....99..638K, 1995ApJ...445..269K} but some cases also report a decoupling between the two \citep{1987ApJ...319..842H, 1990ApJ...359..363G}. The $\textit{Planck}$ dust polarization measurements in 353 GHz allows the precise measurements of polarization direction over all sky giving an insight into the Galactic magnetic fields \citep{2015A&A...576A.104P}. The $\textit{Planck}$ dust polarization results in the regions studied by us show the magnetic fields aligned with the Galactic plane. 
\citet{2011Natur.479..499L} studied six giant molecular cloud complexes in M33 and found a aligned magnetic fields with the   spiral arms of M33 galaxy. This is also consistent to the results of $\lambda~$6$~$cm polarization survey of Galactic plane by \citet{2007A&A...463..993S} who noticed a very uniform large-scale magnetic field running parallel to the Galactic plane. These studies allow to expect a direct correlation between Galactic plane, magnetic field, minor axes and outflow direction unless the turbulence randomize the field orientation. The cloud magnetic field parallel to the galactic plane suggests that the large scale galactic magnetic fields anchor the clouds.

\section{Conclusions}\label{conclude} 
We present the optical polarization study of two cores with embedded low luminosity objects. We have drawn some combined conclusions from our study of magnetic fields in the cores with VeLLOs \citep{2015A&A...573A..34S} in our previous work and low luminosity protostars studied in this work. The main findings of this study are:

\begin{enumerate}
\item The angular offsets between outflow direction and envelope magnetic fields towards L1415 and L1389 are found to be $35\degree$ and $12\degree$, respectively. The mean value of the offset in the cores with similar environments in our previous study (excluding IRAM04191) was $\sim 25\degree$. The sources with relatively less outflow forces are found with lesser offset between envelope magnetic field and outflow directions. This suggests that the outflows from the low luminosity objects do not alter the inherent magnetic field morphology of the cloud and are aligned with the magnetic fields.

\item To estimate the plane-of-the-sky component of the magnetic field strength in the clouds, we have used the CF technique. SF technique has been used to understand the relation between large scale magnetic fields and the turbulent component. The line width information needed for magnetic field strength estimation towards L1415 has been found using the $^{12}$CO(1-0) molecular line observations from TRAO. The plane-of-the-sky component of the magnetic field strength in L1415 and L1389 using CF technique is measured to be 28$\pm15~\mu$G and $\sim149~\mu$G, respectively with a typical uncertainty ($\sigma B_{pos}$) of $\sim0.5B_{pos}$. The plane-of-the-sky magnetic field strength estimated using the SF analysis was found to be 23$~\mu$G and 140$~\mu$G in L1415 and L1389, respectively. The magnetic field strength estimated in L1415 and L1389 are consistent in both CF and SF techniques. At the envelope scale, the dependence of alignment between the outflows and mean field direction, on the magnetic field strength is noticed in this work. 

\item The mean value of angular offset between the envelope magnetic fields and the minor axes of L1415 and L1389 studied in this work is $\sim68\degree$. The mean value of this offset in our previous study of four cores with VeLLOs was found to be $\sim60\degree$. The resultant mean value of the offset in six core becomes $\sim65\degree$. Thus the minor axes of these cores are found to be misaligned with the envelope magnetic fields.

\item The angular offset between the outer magnetic field and inner magnetic field towards L1389 is found to be $14\degree$ suggesting the anchoring of magnetic fields from high density core region to the low density envelope region.

\end{enumerate}

To arrive at a statistically significant conclusion, we have to map the magnetic fields in more number of cores with similar environments. Sub-mm polarization observations in future towards these sources can help us better to understand the correlation between core scale magnetic fields with the envelope fields.

\section{Acknowledgements}
Authors thanks the referee for an encouraging report resulting significant improvement in the paper. The use of SIMBAD and NASA's \textit{SkyView} facility (http://skyview.gsfc.nasa.gov) located at NASA Goddard Space Flight Center is acknowledged. A.S. thanks KASI for providing the post-doctoral research fund. C.W. L. was supported by Basic Science Research Program though the National Research Foundation of Korea (NRF) funded by the Ministry of Education, Science, and Technology (NRF-2016R1A2B4012593). A.S. thanks P. Bhardwaj \& V. Bhagat for their help during the optical observations. A.S. is thankful to the TRAO staff for their continuous  support during the radio observations.

\bibliographystyle{mn2e}
\bibliography{l1415_ref}

%**************************
%\input{table5_pol_res}
\begin{table}
%\begin{small}
\centering
\caption{Polarization results of 224 stars observed in the direction of L1415 and 120 stars observed towards L1389.}\label{tab:polresults}
\begin{tabular}{llllr}\hline
Star  & $\alpha$ (J2000)  & $\delta$ (J2000)  & P $\pm$ $\epsilon_P$ & $\theta$ $\pm$ $\epsilon_{\theta}$  \\ 
 Id  &($\degree$)&($\degree$)& (\%) &($\degree$) \\\hline  
       \hline
       \multicolumn{5}{c}{{\bf L1415}}\\ 
       \hline
       1 &   70.1670&      54.2233&      3.3$\pm$  0.6&   168$\pm$    5  \\
       2 &   70.1901&      54.2042&      3.4$\pm$  0.3&   158$\pm$    2  \\
       3 &   70.2008&      54.1868&      2.0$\pm$  0.3&   154$\pm$    5  \\
       4 &   70.2025&      54.2439&      3.3$\pm$  0.2&   152$\pm$    2  \\
       5 &   70.2117&      54.1901&      2.8$\pm$  0.1&   152$\pm$    0  \\
       6 &   70.2238&      54.3479&      4.0$\pm$  0.6&   159$\pm$    4  \\
       7 &   70.2299&      54.1878&      2.1$\pm$  0.6&   146$\pm$    8  \\
       8 &   70.2351&      54.3925&      6.0$\pm$  1.0&   141$\pm$    5  \\
       9 &   70.2356&      54.2056&      3.6$\pm$  0.2&   150$\pm$    1  \\
      10 &   70.2364&      54.2121&      3.9$\pm$  0.1&   151$\pm$    0  \\
      \multicolumn{5}{c}{}\\
      11 &   70.2376&      54.2705&      4.6$\pm$  0.2&   152$\pm$    1  \\
      12 &   70.2378&      54.3296&      3.8$\pm$  0.4&   139$\pm$    3  \\
      13 &   70.2383&      54.2538&      4.3$\pm$  0.7&   154$\pm$    4  \\
      14 &   70.2423&      54.2291&      3.3$\pm$  0.1&   153$\pm$    1  \\
      15 &   70.2427&      54.3624&      3.5$\pm$  1.3&   155$\pm$   10  \\
      16 &   70.2509&      54.3584&      2.1$\pm$  0.4&   179$\pm$    5  \\
      17 &   70.2519&      54.3844&      4.2$\pm$  0.6&   154$\pm$    4  \\
      18 &   70.2528&      54.2609&      4.0$\pm$  0.1&   150$\pm$    0  \\
      19 &   70.2538&      54.1760&      3.7$\pm$  0.8&   160$\pm$    6  \\
      20 &   70.2563&      54.2748&      2.0$\pm$  0.4&   142$\pm$    6  \\
      \multicolumn{5}{c}{}\\
      21 &   70.2580&      54.3528&      3.6$\pm$  0.4&   151$\pm$    3  \\
      22 &   70.2590&      54.2333&      3.6$\pm$  0.1&   153$\pm$    1  \\
      23 &   70.2645&      54.1940&      3.2$\pm$  0.1&   149$\pm$    1  \\
      24 &   70.2684&      54.2118&      3.2$\pm$  0.4&   146$\pm$    3  \\
      25 &   70.2719&      54.2730&      2.7$\pm$  0.4&   143$\pm$    4  \\
      26 &   70.2722&      54.3548&      3.5$\pm$  0.6&   153$\pm$    5  \\
      27 &   70.2731&      54.3245&      3.5$\pm$  0.6&   168$\pm$    4  \\
      28 &   70.2768&      54.3127&      3.5$\pm$  0.8&   161$\pm$    6  \\
      29 &   70.2786&      54.2239&      2.9$\pm$  0.2&   150$\pm$    1  \\
      30 &   70.2798&      54.3449&      5.2$\pm$  1.1&   155$\pm$    6  \\
      \multicolumn{5}{c}{}\\
      31 &   70.2800&      54.2042&      2.8$\pm$  0.3&   148$\pm$    3  \\
      32 &   70.2803&      54.3630&      6.0$\pm$  0.9&   154$\pm$    4  \\
      33 &   70.2809&      54.2140&      2.5$\pm$  0.8&   147$\pm$    9  \\
      34 &   70.2826&      54.1959&      4.2$\pm$  0.6&   160$\pm$    4  \\
      35 &   70.2836&      54.2461&      2.5$\pm$  0.1&   154$\pm$    2  \\
      36 &   70.2843&      54.1933&      4.3$\pm$  0.6&   151$\pm$    4  \\
      37 &   70.2847&      54.3894&      5.6$\pm$  0.3&   152$\pm$    1  \\
      38 &   70.2874&      54.3880&      4.4$\pm$  1.2&   163$\pm$    7  \\
      39 &   70.2894&      54.2192&      3.4$\pm$  0.6&   147$\pm$    5  \\
      40 &   70.2910&      54.4162&      4.0$\pm$  0.2&   163$\pm$    1  \\
      \multicolumn{5}{c}{}\\
      41 &   70.2923&      54.2490&      3.8$\pm$  0.5&   154$\pm$    3  \\
      42 &   70.2952&      54.2522&      4.0$\pm$  0.2&   150$\pm$    1  \\
      43 &   70.2953&      54.2484&      2.7$\pm$  0.7&   149$\pm$    7  \\
      44 &   70.3018&      54.2512&      2.7$\pm$  0.7&   145$\pm$    7  \\
      45 &   70.3043&      54.5014&      3.0$\pm$  0.6&   166$\pm$    6  \\
      46 &   70.3050&      54.4148&      4.4$\pm$  0.1&   159$\pm$    0  \\
      47 &   70.3084&      54.3108&      3.7$\pm$  0.5&   148$\pm$    4  \\
      48 &   70.3118&      54.2689&      4.0$\pm$  0.4&   147$\pm$    2  \\
      49 &   70.3123&      54.2023&      4.2$\pm$  1.4&   152$\pm$    9  \\
      50 &   70.3136&      54.4179&      5.3$\pm$  1.8&   171$\pm$    9  \\
      \multicolumn{5}{c}{}\\
      51 &   70.3158&      54.4139&      4.5$\pm$  0.9&   160$\pm$    5  \\
      52 &   70.3165&      54.1809&      3.0$\pm$  0.3&   152$\pm$    3  \\
      53 &   70.3203&      54.4078&      1.0$\pm$  0.2&   151$\pm$    7  \\
      54 &   70.3219&      54.1934&      3.7$\pm$  0.6&   153$\pm$    4  \\
      55 &   70.3277&      54.4188&      4.5$\pm$  0.6&   165$\pm$    4  \\
      56 &   70.3291&      54.3568&      3.7$\pm$  0.3&   155$\pm$    2  \\
      \hline
\end{tabular}
%\end{small}
\end{table}
%----------------------------------------------------------------
\begin{table}
%\begin{small}
%\contcaption{}
\begin{tabular}{llllr}\hline
      57 &   70.3343&      54.2485&      2.9$\pm$  0.9&   143$\pm$    8  \\
      58 &   70.3409&      54.1984&      3.9$\pm$  0.3&   155$\pm$    2  \\
      59 &   70.3457&      54.2363&      5.1$\pm$  0.5&   157$\pm$    2  \\
      60 &   70.3507&      54.3522&      4.4$\pm$  0.4&   154$\pm$    2  \\
      \multicolumn{5}{c}{}\\
      61 &   70.3590&      54.5180&      1.2$\pm$  0.5&   172$\pm$   12  \\
      62 &   70.3590&      54.3734&      7.2$\pm$  0.7&   156$\pm$    2  \\
      63 &   70.3619&      54.2404&      3.3$\pm$  0.4&   152$\pm$    3  \\
      64 &   70.3621&      54.5015&      1.9$\pm$  0.3&   167$\pm$    4  \\
      65 &   70.3723&      54.2383&      3.3$\pm$  0.3&   155$\pm$    3  \\
      66 &   70.3762&      54.3372&      0.8$\pm$  0.3&   162$\pm$   11  \\
      67 &   70.3776&      54.5088&      2.1$\pm$  0.3&   173$\pm$    4  \\
      68 &   70.3819&      54.4391&      1.7$\pm$  0.5&   170$\pm$    8  \\
      69 &   70.3877&      54.5447&      1.5$\pm$  0.4&   149$\pm$    7  \\
      70 &   70.4012&      54.5110&      2.2$\pm$  0.3&   152$\pm$    4  \\
      \multicolumn{5}{c}{}\\
      71 &   70.4490&      54.2801&      4.9$\pm$  0.9&   174$\pm$    5  \\
      72 &   70.4525&      54.4072&      3.2$\pm$  0.3&   144$\pm$    3  \\
      73 &   70.4546&      54.5076&      2.6$\pm$  0.1&   154$\pm$    1  \\
      74 &   70.4650&      54.2872&      4.9$\pm$  0.1&   164$\pm$    1  \\
      75 &   70.4653&      54.2880&      5.7$\pm$  2.4&   161$\pm$   12  \\
      76 &   70.4699&      54.4338&      1.6$\pm$  0.7&   146$\pm$   11  \\
      77 &   70.4783&      54.4571&      1.3$\pm$  0.3&   172$\pm$    7  \\
      78 &   70.4814&      54.2858&      5.1$\pm$  0.6&   157$\pm$    3  \\
      79 &   70.4819&      54.2186&      4.0$\pm$  0.3&   153$\pm$    2  \\
      80 &   70.4839&      54.2579&      5.7$\pm$  1.2&   155$\pm$    6  \\
      \multicolumn{5}{c}{}\\
      81 &   70.4861&      54.4030&      2.7$\pm$  0.9&   139$\pm$    9  \\
      82 &   70.4861&      54.4030&      2.7$\pm$  0.9&   139$\pm$    9  \\
      83 &   70.4867&      54.4050&      2.1$\pm$  0.1&   139$\pm$    1  \\
      84 &   70.4916&      54.2647&      3.6$\pm$  1.6&   167$\pm$   12  \\
      85 &   70.4930&      54.2799&      4.6$\pm$  0.7&   156$\pm$    4  \\
      86 &   70.4941&      54.4454&      1.9$\pm$  0.8&   171$\pm$   12  \\
      87 &   70.4951&      54.4455&      1.9$\pm$  0.4&   174$\pm$    5  \\
      88 &   70.4956&      54.4306&      1.0$\pm$  0.3&   175$\pm$    8  \\
      89 &   70.4964&      54.2582&      1.3$\pm$  0.2&   172$\pm$    4  \\
      90 &   70.4974&      54.4034&      2.5$\pm$  0.7&   144$\pm$    8  \\
      91 &   70.4984&      54.2376&      5.8$\pm$  0.9&   150$\pm$    4  \\
      92 &   70.5006&      54.2579&      4.3$\pm$  0.8&   149$\pm$    5  \\
      93 &   70.5096&      54.4286&      2.9$\pm$  0.1&   167$\pm$    1  \\
      94 &   70.5151&      54.2370&      4.3$\pm$  1.3&   152$\pm$    8  \\
      95 &   70.5215&      54.4471&      0.8$\pm$  0.2&   179$\pm$    8  \\
      96 &   70.5244&      54.2413&      1.0$\pm$  0.3&   175$\pm$    7  \\
      97 &   70.5251&      54.3035&      2.4$\pm$  0.6&   145$\pm$    7  \\
      98 &   70.5265&      54.3793&      5.0$\pm$  0.6&   148$\pm$    3  \\
      99 &   70.5282&      54.2902&      3.0$\pm$  0.4&   158$\pm$    4  \\
     100 &   70.5287&      54.4221&      3.5$\pm$  0.6&   156$\pm$    5  \\
     \multicolumn{5}{c}{}\\
     101 &   70.5289&      54.4220&      1.9$\pm$  0.8&   174$\pm$   11  \\
     102 &   70.5307&      54.4046&      1.9$\pm$  0.4&   154$\pm$    6  \\
     103 &   70.5309&      54.4046&      2.2$\pm$  0.6&   153$\pm$    7  \\
     104 &   70.5322&      54.4055&      1.8$\pm$  0.4&   158$\pm$    6  \\
     105 &   70.5362&      54.3861&      5.4$\pm$  0.5&   176$\pm$    3  \\
     106 &   70.5363&      54.3860&      4.1$\pm$  0.8&   152$\pm$    5  \\
     107 &   70.5374&      54.4618&      2.4$\pm$  1.1&   165$\pm$   12  \\
     108 &   70.5403&      54.2557&      1.8$\pm$  0.3&   155$\pm$    5  \\
     109 &   70.5506&      54.4335&      3.2$\pm$  0.7&   153$\pm$    6  \\
     110 &   70.5508&      54.4335&      2.5$\pm$  0.7&   172$\pm$    7  \\
     \multicolumn{5}{c}{}\\
     111 &   70.5513&      54.3673&      4.4$\pm$  0.4&   155$\pm$    3  \\
     112 &   70.5513&      54.3673&      4.4$\pm$  0.4&   155$\pm$    3  \\
     \hline
\end{tabular}
%\end{small}
\end{table}
%----------------------------------------------------------------
\begin{table}
%\begin{small}
%\contcaption{}
\begin{tabular}{llllr}\hline
     113 &   70.5518&      54.2333&      3.4$\pm$  0.2&   151$\pm$    2  \\
     114 &   70.5520&      54.4388&      3.6$\pm$  0.9&   145$\pm$    7  \\
     115 &   70.5525&      54.4832&      1.9$\pm$  0.1&   161$\pm$    1  \\
     116 &   70.5543&      54.2064&      7.4$\pm$  0.8&   144$\pm$    3  \\
     117 &   70.5553&      54.2148&      2.4$\pm$  0.3&   133$\pm$    3  \\
     118 &   70.5557&      54.3012&      2.4$\pm$  0.8&   161$\pm$    9  \\
     119 &   70.5560&      54.2295&      3.9$\pm$  0.9&   147$\pm$    6  \\
     120 &   70.5564&      54.2446&      4.3$\pm$  1.1&   157$\pm$    7  \\
     \multicolumn{5}{c}{}\\
     121 &   70.5596&      54.4184&      3.0$\pm$  0.2&   166$\pm$    2  \\
     122 &   70.5598&      54.4183&      3.0$\pm$  0.2&   170$\pm$    2  \\
     123 &   70.5599&      54.4183&      3.3$\pm$  0.2&   171$\pm$    1  \\
     124 &   70.5601&      54.2201&      2.1$\pm$  0.4&   135$\pm$    5  \\
     125 &   70.5613&      54.2546&      3.0$\pm$  0.9&   161$\pm$    8  \\
     126 &   70.5685&      54.2146&      1.6$\pm$  0.1&   147$\pm$    1  \\
     127 &   70.5706&      54.4824&      1.1$\pm$  0.2&   167$\pm$    5  \\
     128 &   70.5746&      54.2202&      1.3$\pm$  0.3&   133$\pm$    7  \\
     129 &   70.5768&      54.4388&      1.6$\pm$  0.3&   175$\pm$    6  \\
     130 &   70.5768&      54.4389&      1.1$\pm$  0.2&   179$\pm$    6  \\
     \multicolumn{5}{c}{}\\
     131 &   70.5768&      54.4129&      5.2$\pm$  0.9&   166$\pm$    4  \\
     132 &   70.5775&      54.2526&      2.1$\pm$  1.0&   161$\pm$   12  \\
     133 &   70.5804&      54.2515&      2.9$\pm$  0.4&   159$\pm$    4  \\
     134 &   70.5870&      54.2569&      2.9$\pm$  1.3&   145$\pm$   12  \\
     135 &   70.5878&      54.4501&      0.7$\pm$  0.2&   174$\pm$    8  \\
     136 &   70.5890&      54.4078&      3.2$\pm$  0.5&   158$\pm$    5  \\
     137 &   70.5891&      54.4077&      3.0$\pm$  0.8&   171$\pm$    7  \\
     138 &   70.5891&      54.4077&      2.9$\pm$  0.6&   171$\pm$    5  \\
     139 &   70.5892&      54.2546&      3.1$\pm$  0.4&   152$\pm$    4  \\
     140 &   70.5895&      54.4752&      1.5$\pm$  0.7&   173$\pm$   12  \\
     \multicolumn{5}{c}{}\\
     141 &   70.5898&      54.4727&      3.2$\pm$  0.6&   172$\pm$    5  \\
     142 &   70.5924&      54.2316&      2.1$\pm$  0.5&   126$\pm$    6  \\
     143 &   70.5944&      54.4206&      3.2$\pm$  0.6&   164$\pm$    6  \\
     144 &   70.5945&      54.4205&      3.3$\pm$  0.6&   174$\pm$    5  \\
     145 &   70.5946&      54.4206&      3.3$\pm$  0.5&   169$\pm$    4  \\
     146 &   70.5961&      54.3854&      2.2$\pm$  0.3&   160$\pm$    4  \\
     147 &   70.5961&      54.3854&      2.1$\pm$  0.2&   157$\pm$    3  \\
     148 &   70.5968&      54.2469&      1.2$\pm$  0.2&   158$\pm$    5  \\
     149 &   70.5978&      54.4714&      1.0$\pm$  0.3&   179$\pm$    8  \\
     150 &   70.5984&      54.4010&      3.1$\pm$  0.8&   166$\pm$    6  \\
     \multicolumn{5}{c}{}\\
     151 &   70.5985&      54.4010&      2.9$\pm$  1.0&   145$\pm$    9  \\
     152 &   70.5985&      54.4011&      4.0$\pm$  1.0&   169$\pm$    8  \\
     153 &   70.5991&      54.2382&      2.8$\pm$  0.5&   142$\pm$    5  \\
     154 &   70.5999&      54.2878&      4.7$\pm$  0.9&   156$\pm$    5  \\
     155 &   70.6018&      54.4323&      3.4$\pm$  0.4&   158$\pm$    3  \\
     156 &   70.6019&      54.4322&      3.1$\pm$  0.3&   162$\pm$    3  \\
     157 &   70.6046&      54.4876&      0.8$\pm$  0.1&   155$\pm$    5  \\
     158 &   70.6050&      54.4142&      2.5$\pm$  0.1&   153$\pm$    2  \\
     159 &   70.6050&      54.4142&      2.3$\pm$  0.1&   160$\pm$    1  \\
     160 &   70.6050&      54.4142&      1.9$\pm$  0.1&   164$\pm$    2  \\
     \multicolumn{5}{c}{}\\
     161 &   70.6111&      54.3892&      1.6$\pm$  0.3&   150$\pm$    5  \\
     162 &   70.6111&      54.3892&      1.7$\pm$  0.3&   148$\pm$    5  \\
     163 &   70.6111&      54.3892&      1.5$\pm$  0.3&   153$\pm$    6  \\
     164 &   70.6114&      54.2784&      1.9$\pm$  0.5&   150$\pm$    7  \\
     165 &   70.6149&      54.2324&      1.6$\pm$  0.3&   148$\pm$    5  \\
     166 &   70.6173&      54.4155&      3.0$\pm$  0.4&   151$\pm$    3  \\
     167 &   70.6174&      54.4155&      2.6$\pm$  0.6&   154$\pm$    6  \\
     168 &   70.6209&      54.3918&      3.1$\pm$  0.5&   165$\pm$    4  \\
          \hline
\end{tabular}
%\end{small}
\end{table}
%----------------------------------------------------------------
\begin{table}
%\begin{small}
%\contcaption{}
\begin{tabular}{llllr}\hline
     169 &   70.6209&      54.3919&      2.3$\pm$  0.6&   161$\pm$    8  \\
     170 &   70.6212&      54.4939&      1.1$\pm$  0.4&   156$\pm$    9  \\
     \multicolumn{5}{c}{}\\
     171 &   70.6258&      54.4067&      4.6$\pm$  1.5&   152$\pm$    9  \\
     172 &   70.6284&      54.3806&      2.3$\pm$  0.7&   133$\pm$    8  \\
     173 &   70.6334&      54.4058&      2.5$\pm$  1.0&   162$\pm$   10  \\
     174 &   70.6340&      54.4247&      3.4$\pm$  1.1&   146$\pm$    9  \\
     175 &   70.6376&      54.4888&      1.2$\pm$  0.3&   167$\pm$    7  \\
     176 &   70.6390&      54.2297&      1.5$\pm$  0.6&   124$\pm$   10  \\
     177 &   70.6400&      54.4911&      1.6$\pm$  0.4&   163$\pm$    6  \\
     178 &   70.6406&      54.3840&      3.2$\pm$  0.4&   135$\pm$    3  \\
     179 &   70.6406&      54.3840&      3.1$\pm$  0.4&   127$\pm$    3  \\
     180 &   70.6506&      54.5006&      2.2$\pm$  0.8&   157$\pm$   10  \\
     \multicolumn{5}{c}{}\\
     181 &   70.6571&      54.4084&      4.1$\pm$  1.1&   160$\pm$    7  \\
     182 &   70.6684&      54.4058&      4.5$\pm$  1.1&   153$\pm$    7  \\
     183 &   70.6684&      54.4058&      5.9$\pm$  1.1&   133$\pm$    4  \\
     184 &   70.6726&      54.4609&      1.6$\pm$  0.1&   157$\pm$    1  \\
     185 &   70.6765&      54.4570&      1.3$\pm$  0.4&   166$\pm$    8  \\
     186 &   70.6772&      54.4604&      2.3$\pm$  0.4&   155$\pm$    4  \\
     187 &   70.6780&      54.4636&      2.1$\pm$  0.3&   169$\pm$    4  \\
     188 &   70.6794&      54.3912&      3.9$\pm$  0.7&   152$\pm$    5  \\
     189 &   70.6806&      54.3941&      2.7$\pm$  0.3&   135$\pm$    2  \\
     190 &   70.6828&      54.4090&      1.2$\pm$  0.1&   176$\pm$    2  \\
     \multicolumn{5}{c}{}\\
     191 &   70.6828&      54.4089&      0.9$\pm$  0.1&   159$\pm$    3  \\
     192 &   70.6844&      54.2896&      4.4$\pm$  0.1&   157$\pm$    0  \\
     193 &   70.6898&      54.4716&      4.0$\pm$  0.8&   147$\pm$    6  \\
     194 &   70.6899&      54.4804&      1.0$\pm$  0.5&   145$\pm$   13  \\
     195 &   70.6916&      54.4223&      1.4$\pm$  0.7&   160$\pm$   12  \\
     196 &   70.6977&      54.4736&      1.6$\pm$  0.3&   147$\pm$    6  \\
     197 &   70.6989&      54.3668&      0.7$\pm$  0.2&   158$\pm$    7  \\
     198 &   70.7005&      54.4393&      2.3$\pm$  0.5&   149$\pm$    6  \\
     199 &   70.7014&      54.3353&      2.6$\pm$  0.2&   169$\pm$    2  \\
     200 &   70.7026&      54.4625&      3.2$\pm$  1.2&   160$\pm$   10  \\
     \multicolumn{5}{c}{}\\
     201 &   70.7060&      54.3204&      2.6$\pm$  0.1&   157$\pm$    1  \\
     202 &   70.7141&      54.2462&      2.9$\pm$  0.7&   149$\pm$    6  \\
     203 &   70.7263&      54.3366&      0.7$\pm$  0.1&   165$\pm$    4  \\
     204 &   70.7283&      54.2612&      3.8$\pm$  1.4&   153$\pm$   10  \\
     205 &   70.7288&      54.4292&      2.6$\pm$  1.1&   169$\pm$   11  \\
     206 &   70.7340&      54.3414&      1.8$\pm$  0.8&   149$\pm$   13  \\
     207 &   70.7367&      54.3438&      2.3$\pm$  0.7&   133$\pm$    8  \\
     208 &   70.7394&      54.2847&      4.5$\pm$  0.2&   153$\pm$    1  \\
     209 &   70.7395&      54.4522&      0.8$\pm$  0.1&   166$\pm$    2  \\
     210 &   70.7428&      54.2850&      4.7$\pm$  0.3&   150$\pm$    1  \\
     \multicolumn{5}{c}{}\\
     211 &   70.7515&      54.2677&      4.5$\pm$  1.2&   140$\pm$    7  \\
     212 &   70.7527&      54.2716&      5.1$\pm$  1.1&   150$\pm$    6  \\
     213 &   70.7606&      54.2624&      0.8$\pm$  0.2&   174$\pm$    6  \\
     214 &   70.7610&      54.3067&      3.3$\pm$  1.6&   156$\pm$   13  \\
     215 &   70.7747&      54.2369&      2.7$\pm$  0.5&   149$\pm$    5  \\
     216 &   70.7762&      54.2649&      4.3$\pm$  0.4&   147$\pm$    3  \\
     217 &   70.7783&      54.3027&      4.0$\pm$  0.9&   153$\pm$    6  \\
     218 &   70.7816&      54.2471&      2.7$\pm$  0.3&   144$\pm$    3  \\
     219 &   70.7924&      54.2499&      3.4$\pm$  0.3&   142$\pm$    3  \\
     220 &   70.8041&      54.2840&      3.4$\pm$  0.4&   153$\pm$    3  \\
     221 &   70.8043&      54.2788&      3.2$\pm$  0.5&   148$\pm$    4  \\
     222 &   70.8062&      54.2762&      3.0$\pm$  1.0&   156$\pm$    9  \\
     223 &   70.8280&      54.2748&      3.9$\pm$  0.8&   143$\pm$    5  \\
     224 &   70.8511&      54.2709&      1.6$\pm$  0.5&   125$\pm$    8  \\
     \hline
\end{tabular}
%\end{small}
\end{table}
%----------------------------------------------------------------
\begin{table}
%\begin{small}
%\contcaption{}
\begin{tabular}{llllr}
     \hline
     \multicolumn{5}{c}{{\bf L1389}}\\
     \hline
      1 &      61.0506 &     56.9603&      4.0$\pm$     0.7 &    134$\pm$     5  \\
      2 &      61.0506 &     56.9603&      4.0$\pm$     0.7 &    134$\pm$     5	 \\
      3 &      61.0508 &     56.9344&      4.6$\pm$     0.3 &    136$\pm$     2	 \\
      4 &      61.0621 &     56.9307&      1.3$\pm$     0.3 &    118$\pm$     6	 \\
      5 &      61.0655 &     56.9464&      1.0$\pm$     0.2 &    136$\pm$     7	 \\
      6 &      61.0763 &     56.9219&      3.8$\pm$     1.8 &    148$\pm$    13	 \\
      7 &      61.0820 &     56.9500&      5.1$\pm$     1.4 &    138$\pm$     7	 \\
      8 &      61.0865 &     56.9199&      3.7$\pm$     0.9 &    128$\pm$     7	 \\
      9 &      61.0898 &     56.9437&      3.4$\pm$     0.5 &    137$\pm$     4	 \\
     10 &      61.0935 &     57.1189&      3.8$\pm$     0.4 &    121$\pm$     3	 \\
     \multicolumn{5}{c}{}\\
     11 &      61.0946 &     56.9167&      4.3$\pm$     1.1 &    142$\pm$     7	 \\
     12 &      61.1010 &     56.9515&      3.8$\pm$     0.4 &    137$\pm$     3	 \\
     13 &      61.1061 &     56.9814&      3.3$\pm$     1.2 &    137$\pm$    10	 \\
     14 &      61.1101 &     57.1303&      3.5$\pm$     0.5 &    129$\pm$     4	 \\
     15 &      61.1129 &     56.9169&      4.3$\pm$     0.6 &    140$\pm$     4	 \\
     16 &      61.1149 &     56.9422&      4.2$\pm$     0.2 &    139$\pm$     1	 \\
     17 &      61.1161 &     56.9969&      4.0$\pm$     1.0 &    133$\pm$     7	 \\
     18 &      61.1201 &     56.9134&      2.1$\pm$     0.1 &    136$\pm$     1	 \\
     19 &      61.1232 &     56.9160&      5.0$\pm$     1.2 &    143$\pm$     7	 \\
     20 &      61.1313 &     57.1542&      3.3$\pm$     0.1 &    136$\pm$     1	 \\
     \multicolumn{5}{c}{}\\
     21 &      61.1314 &     56.9178&      5.1$\pm$     0.6 &    129$\pm$     3	 \\
     22 &      61.1390 &     57.1671&      2.5$\pm$     0.3 &    135$\pm$     3	 \\
     23 &      61.1391 &     56.9627&      3.6$\pm$     0.3 &    134$\pm$     2	 \\
     24 &      61.1394 &     57.1184&      3.8$\pm$     0.4 &    139$\pm$     3	 \\
     25 &      61.1397 &     56.9210&      4.7$\pm$     1.8 &    127$\pm$    10	 \\
     26 &      61.1431 &     56.8963&      4.5$\pm$     1.4 &    144$\pm$     8	 \\
     27 &      61.1434 &     56.8906&      4.0$\pm$     1.8 &    137$\pm$    12	 \\
     28 &      61.1455 &     57.1118&      3.5$\pm$     0.1 &    134$\pm$     1	 \\
     29 &      61.1471 &     57.1366&      4.1$\pm$     0.3 &    134$\pm$     2	 \\
     30 &      61.1480 &     57.0080&      3.9$\pm$     0.5 &    140$\pm$     4	 \\
     \multicolumn{5}{c}{}\\
     31 &      61.1493 &     56.9722&      5.3$\pm$     0.4 &    139$\pm$     2	 \\
     32 &      61.1517 &     57.0154&      3.2$\pm$     0.6 &    142$\pm$     5	 \\
     33 &      61.1521 &     57.0084&      3.7$\pm$     0.6 &    136$\pm$     4	 \\
     34 &      61.1526 &     57.1845&      3.7$\pm$     0.1 &    137$\pm$     1	 \\
     35 &      61.1543 &     57.0066&      4.4$\pm$     0.4 &    138$\pm$     2	 \\
     36 &      61.1584 &     56.9707&      2.8$\pm$     0.8 &    144$\pm$     8	\\
     37 &      61.1639 &     56.8965&      3.1$\pm$     0.8 &    145$\pm$     7	 \\
     38 &      61.1640 &     57.1287&      2.8$\pm$     0.8 &    135$\pm$     7	 \\
     39 &      61.1659 &     57.1021&      3.0$\pm$     0.5 &    129$\pm$     4	 \\
     40 &      61.1677 &     57.1293&      3.3$\pm$     0.5 &    136$\pm$     4	 \\
     \multicolumn{5}{c}{}\\
     41 &      61.1766 &     56.9995&      3.2$\pm$     0.3 &    157$\pm$     3	 \\
     42 &      61.1807 &     57.1193&      3.6$\pm$     0.5 &    132$\pm$     4	 \\
     43 &      61.1816 &     57.1579&      3.8$\pm$     0.2 &    133$\pm$     1	 \\
     44 &      61.1864 &     57.0017&      1.5$\pm$     0.4 &    133$\pm$     7	 \\
     45 &      61.1876 &     57.0271&      3.7$\pm$     0.5 &    137$\pm$     4	 \\
     46 &      61.1976 &     57.1483&      4.6$\pm$     0.6 &    131$\pm$     3	 \\
     47 &      61.2021 &     56.9997&      0.7$\pm$     0.1 &    129$\pm$     4	 \\
     48 &      61.2063 &     57.0914&      3.4$\pm$     0.1 &    132$\pm$     1	 \\
     49 &      61.2119 &     57.0353&      3.4$\pm$     0.6 &    144$\pm$     5	 \\
     50 &      61.2125 &     57.0312&      2.3$\pm$     0.5 &    138$\pm$     6	 \\
     \multicolumn{5}{c}{}\\
     51 &      61.2127 &     57.0916&      2.7$\pm$     0.1 &    135$\pm$     1	 \\
     52 &      61.2129 &     57.0309&      2.3$\pm$     0.2 &    138$\pm$     3	 \\
     53 &      61.2137 &     57.0913&      4.0$\pm$     0.3 &    133$\pm$     2	 \\
     54 &      61.2154 &     56.9815&      2.3$\pm$     0.2 &    135$\pm$     3	 \\
     55 &      61.2156 &     56.9333&      4.0$\pm$     1.6 &    139$\pm$    11	 \\
        \hline
\end{tabular}
%\end{small}
\end{table}
%----------------------------------------------------------------
\begin{table}
%\begin{small}
%\contcaption{}
\begin{tabular}{llllr}\hline
     56 &      61.2176 &     57.1152&      3.3$\pm$     0.5 &    135$\pm$     4	 \\
     57 &      61.2186 &     57.1151&      2.9$\pm$     0.4 &    132$\pm$     4	 \\
     58 &      61.2202 &     57.1592&      3.0$\pm$     0.6 &    132$\pm$     6	 \\
     59 &      61.2228 &     57.1960&      3.1$\pm$     0.4 &    140$\pm$     3	 \\
     60 &      61.2234 &     57.0616&      3.7$\pm$     0.1 &    134$\pm$     2	 \\
     \multicolumn{5}{c}{}\\
     61 &      61.2237 &     56.9011&      2.5$\pm$     1.1 &    126$\pm$    11	 \\
     62 &      61.2239 &     57.0236&      1.5$\pm$     0.5 &    130$\pm$     9	 \\
     63 &      61.2249 &     57.1955&      2.2$\pm$     0.3 &    135$\pm$     4	 \\
     64 &      61.2256 &     57.1185&      3.9$\pm$     0.1 &    130$\pm$     1	 \\
     65 &      61.2265 &     57.1183&      4.1$\pm$     0.1 &    129$\pm$     1	 \\
     66 &      61.2267 &     57.0693&      3.4$\pm$     0.1 &    134$\pm$     1	 \\
     67 &      61.2270 &     57.0689&      3.7$\pm$     0.1 &    134$\pm$     1	 \\
     68 &      61.2295 &     57.0467&      4.7$\pm$     0.9 &    135$\pm$     5	 \\
     69 &      61.2310 &     57.1506&      4.6$\pm$     0.4 &    130$\pm$     2	 \\
     70 &      61.2310 &     57.0006&      3.3$\pm$     0.1 &    135$\pm$     1	 \\
     \multicolumn{5}{c}{}\\
     71 &      61.2318 &     56.9568&      3.2$\pm$     1.3 &    133$\pm$    11	 \\
     72 &      61.2344 &     56.9201&      2.8$\pm$     0.2 &    137$\pm$     2	 \\
     73 &      61.2347 &     56.9982&      3.8$\pm$     0.4 &    137$\pm$     3	 \\
     74 &      61.2381 &     56.9792&      3.2$\pm$     0.4 &    134$\pm$     3	 \\
     75 &      61.2391 &     57.1492&      2.6$\pm$     0.3 &    137$\pm$     3	 \\
     76 &      61.2407 &     57.0363&      3.2$\pm$     0.3 &    145$\pm$     3	 \\
     77 &      61.2431 &     57.1453&      2.7$\pm$     0.1 &    137$\pm$     1	 \\
     78 &      61.2438 &     57.0269&      4.1$\pm$     0.6 &    132$\pm$     4	 \\
     79 &      61.2462 &     57.1802&      3.8$\pm$     0.4 &    134$\pm$     3	 \\
     80 &      61.2479 &     57.0891&      2.4$\pm$     0.1 &    134$\pm$     1	 \\
     \multicolumn{5}{c}{}\\
     81 &      61.2521 &     57.0302&      3.5$\pm$     0.4 &    131$\pm$     3	 \\
     82 &      61.2569 &     56.9554&      3.1$\pm$     0.7 &    141$\pm$     6	 \\
     83 &      61.2617 &     57.0785&      3.1$\pm$     0.2 &    136$\pm$     2	 \\
     84 &      61.2624 &     57.1660&      4.1$\pm$     0.9 &    126$\pm$     6	 \\
     85 &      61.2645 &     57.1067&      4.3$\pm$     0.4 &    135$\pm$     2	 \\
     86 &      61.2655 &     57.1065&      4.1$\pm$     0.4 &    134$\pm$     2	 \\
     87 &      61.2663 &     57.0791&      3.7$\pm$     0.1 &    138$\pm$     1	 \\
     88 &      61.2667 &     57.1496&      5.0$\pm$     0.6 &    128$\pm$     3	 \\
     89 &      61.2762 &     56.9743&      3.1$\pm$     0.1 &    137$\pm$     1	 \\
     90 &      61.2773 &     57.0502&      3.1$\pm$     0.7 &    128$\pm$     7	 \\
     \multicolumn{5}{c}{}\\
     91 &      61.2785 &     57.1301&      3.2$\pm$     0.1 &    136$\pm$     3	 \\
     92 &      61.2796 &     57.0699&      3.0$\pm$     0.2 &    143$\pm$     2	 \\
     93 &      61.2798 &     57.0334&      3.8$\pm$     0.4 &    135$\pm$     3	 \\
     94 &      61.2799 &     57.0695&      3.3$\pm$     0.2 &    141$\pm$     2	 \\
     95 &      61.2800 &     57.1327&      3.0$\pm$     0.1 &    140$\pm$     1	 \\
     96 &      61.2801 &     57.0331&      2.6$\pm$     0.4 &    132$\pm$     5	 \\
     97 &      61.2812 &     56.9871&      3.1$\pm$     0.1 &    143$\pm$     1	 \\
     98 &      61.2886 &     57.0916&      2.8$\pm$     0.3 &    137$\pm$     3	 \\
     99 &      61.2915 &     57.0906&      2.5$\pm$     0.4 &    133$\pm$     4	 \\
    100 &      61.2969 &     57.1221&      1.8$\pm$     0.2 &    123$\pm$     3	 \\
     \multicolumn{5}{c}{}\\
    101 &      61.2979 &     57.1219&      1.9$\pm$     0.2 &    129$\pm$     2	 \\
    102 &      61.3064 &     56.9901&      2.7$\pm$     0.3 &    130$\pm$     4	 \\
    103 &      61.3095 &     57.0807&      2.3$\pm$     0.6 &    124$\pm$     7	 \\
    104 &      61.3140 &     57.0398&      2.6$\pm$     0.2 &    133$\pm$     2	 \\
    105 &      61.3144 &     57.0394&      2.0$\pm$     0.2 &    130$\pm$     3	 \\
    106 &      61.3170 &     57.0380&      3.7$\pm$     0.4 &    142$\pm$     2	 \\
    107 &      61.3176 &     57.0317&      3.1$\pm$     0.7 &    141$\pm$     6	 \\
    108 &      61.3230 &     57.0879&      2.1$\pm$     0.1 &    138$\pm$     1	 \\
    109 &      61.3266 &     56.9906&      2.3$\pm$     0.6 &    142$\pm$     7	 \\
    110 &      61.3327 &     57.0512&      3.1$\pm$     0.3 &    138$\pm$     2	 \\
    \multicolumn{5}{c}{}\\
    111 &      61.3327 &     56.9901&      2.2$\pm$     0.2 &    138$\pm$     3	 \\
        \hline
\end{tabular}
%\end{small}
\end{table}
%----------------------------------------------------------------
\begin{table}
%\begin{small}
%\contcaption{}
\begin{tabular}{llllr}\hline
    112 &      61.3330 &     57.0508&      3.3$\pm$     0.3 &    133$\pm$     2	 \\
    113 &      61.3352 &     56.9933&      4.3$\pm$     0.8 &    141$\pm$     5	 \\
    114 &      61.3359 &     56.9721&      3.5$\pm$     0.4 &    123$\pm$     3	 \\
    115 &      61.3419 &     57.0655&      2.5$\pm$     0.2 &    132$\pm$     2	 \\
    116 &      61.3492 &     57.0346&      3.8$\pm$     0.3 &    133$\pm$     2	 \\
    117 &      61.3513 &     57.0036&      2.4$\pm$     0.2 &    131$\pm$     2	 \\
    118 &      61.3559 &     57.0818&      3.5$\pm$     0.3 &    132$\pm$     2	 \\
    119 &      61.3609 &     57.0704&      2.3$\pm$     0.5 &    148$\pm$     7	 \\
    120 &      61.3687 &     57.0909&      1.4$\pm$     0.1 &    137$\pm$     2	 \\
\hline
\hline
\end{tabular}
%\end{small}
\end{table}
%************************
\end{document}